\documentclass[aps,prx,twocolumn,groupedaddress,longbibliography]{revtex4-1}

\usepackage{graphicx}
\usepackage{amssymb}
\usepackage{amsmath}
\usepackage[markup=underlined]{changes}

\newcommand{\xv}{{\bf x}}

\newcommand{\uv}{{\bf u}}

\newcommand{\tv}{{\bf t}}

\newcommand{\grad}{{\bf \nabla}}

\begin{document}


\title{Defect-driven shape instabilities of bundles}


\author{Isaac R. Bruss}
\affiliation{School of Engineering and Applied Sciences, Harvard University, Cambridge, Massachusetts 01451, USA}
\author{Gregory M. Grason}
\affiliation{Department of Polymer Science and Engineering, University of Massachusetts, Amherst, Massachusetts 01003, USA}


\date{\today}

\begin{abstract}
Topological defects are crucial to the thermodynamics and structure of condensed matter systems. For instance, when incorporated into crystalline membranes like graphene, disclinations with positive and negative topological charge elastically buckle the material into conical and saddle-like shapes respectively. A recently uncovered mapping between the inter-element spacing in 2D columnar structures and the metric properties of curved surfaces motivates basic questions about the interplay between defects in the cross section of a columnar bundle and its 3D shape. Such questions are critical to the structure of a broad class of filamentous materials, from biological assemblies like protein fibers to nano- or micro-structured synthetic materials like carbon nanotube bundles. Here, we explore the buckling behavior for elementary disclinations in hexagonal bundles using a combination of continuum elasticity theory and numerical simulations of discrete-filaments. We show that shape instabilities are controlled by a single material-dependent parameter that characterizes the ratio of inter-filament to intra-filament elastic energies. Along with a host of previously unknown shape equilibria---the filamentous analogs to the conical and saddle-like shapes of defective membranes---we find a profoundly asymmetric response to positive and negative topologically charged defects in the infinite length limit that is without parallel to the membrane analog. The highly non-linear dependence on the sign of the disclination charge is shown to have a purely geometric origin, stemming the from the distinct compatibility (or incompatibility) of effectively positive- (or negative-) curvature geometries with lengthwise-constant filament spacing.
\end{abstract}


\maketitle

\section{Introduction}
\label{section:intro}
Topological defects are fundamental to the properties of ordered materials, from their structure and thermodynamics to their dynamics and mechanical response. There is long history, dating back to some of the earliest mathematical models of defects \cite{Kondo1964, Kroner1980, Bilby1955}, of understanding the non-linear influence of topological defects on material structure through the lens of differential geometry. In such a description, topological defects are understood as sources for metric deformation in solid media \cite{Moshe2015}, specifically curvature multipoles, leading to intrinsic stresses that reshape the material and its stress response.   Far more than descriptive, the relationship between intrinsic (Gaussian) curvature and topological defects, becomes even more profound for ordered systems that are free to ``reshape their metrics'', such as 2D ordered membranes of both crystalline and liquid crystalline (e.g.\ 2D in-plane polar, nematic or smectic order) varieties \cite{refId0, Seung1988a, refId1, Bowick2008a, Hirst2013a}. The flexibility of out-of-plane deformations in such systems gives rise to an instability, in which the in-plane metric of a thin membrane may adapt to its 3D environment to accommodate the non-Euclidean geometry favored by disclinations (and multi-pole combinations thereof, such as dislocations), leading to a spontaneous buckling of sufficiently thin membranes~\cite{Seung1988a}. Beyond the relevance to self-organized matter, the mechanical and geometric principles of defects in elastic sheets is now a cornerstone of the current approaches for designing and engineering 2D origami and kirigami materials \cite{Silverberg2015, Blees2015a, Grosso2015, Castle2014a}.

In this paper, we analyze the defect-induced geometric instability of a parallel class of 2D ordered matter: columnar or filamentous bundles. This is done using a structural model that describes cohesive assemblies of many quasi-1D elements (e.g.~filaments or columns) possessing 2D order transverse to their backbone. This model applies to a broad class of materials, from assemblies of flexible filaments (e.g.~protein filaments \cite{Fratzl2003, Fratzl2008, Hulmes1995a, Weisel1987, Parry2005} or synthetic nanotubes \cite{Journet1997a, Rols2000, Zhang2002, Zhang2004}) that self-assembled into cable-like structures via attractive interactions between fibers of self-stacking molecules \cite{Brunsveld2001, Douglas2009}, to finite domains of columnar liquid crystals \cite{Livolant2000, Tortora2011, Verhoeff2012}, or even condensed phases of vortices on multicomponent superconductors \cite{Babaev2012}. This class of material is two-dimensionally ordered in the sense that it retains translational symmetry along the filaments or columns \cite{Chaikin2000}. Recent works \cite{grason2010a, bruss2012c, grason2015a} demonstrate that columnar systems, like membranes, are also capable of altering their geometry by modifying their 3D embedding. But unlike membranes which deflect out-of-plane, this deformation occurs through the geometrically non-linear coupling between the columns' orientations and transverse spacings (i.e.\ their metric). A well known example of such a coupling in bulk columnar media is the Helfrich-Hurault instability \cite{Oswald2005}, in which a sufficiently large transverse tension drives a non-uniform tilt of the columns, thus maintaining a more uniform inter-column spacing at the expense of bending \cite{Jonathan1991}. Driven by the mechanics of metric deformation, this instability is the columnar analogy to the Euler buckling of a 2D elastic sheet under compression, which underlies the 3D buckling behavior of defective crystalline membranes. In contrast, for columnar media with defects, the consequences of the geometric instability to tension are not known.

\begin{figure}[h]
\centering
\includegraphics[width=0.48\textwidth]{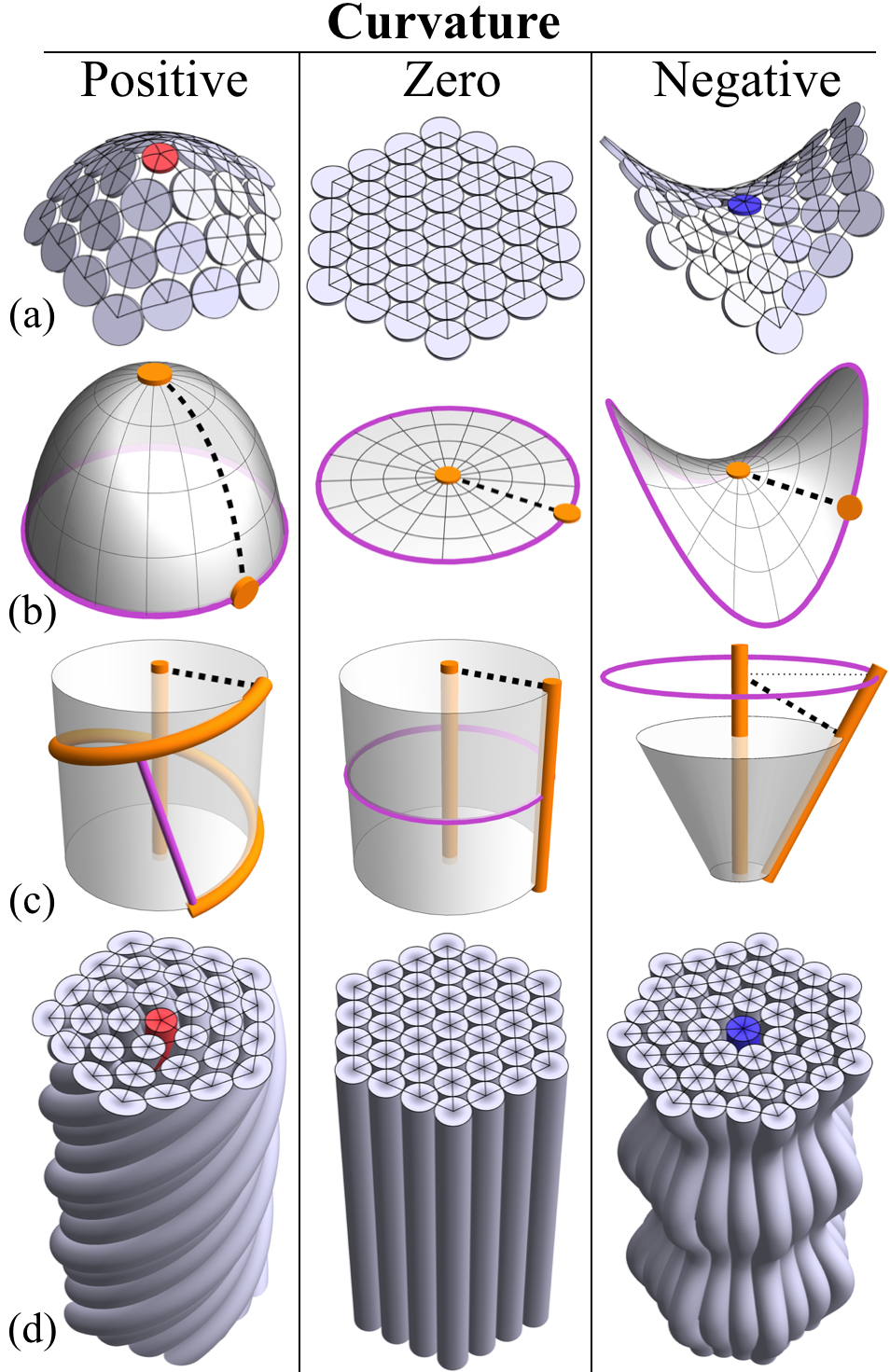}
\caption{(a) A flat flexible crystalline membrane will buckle into a positively curved dome (left) or negatively curved saddle (right) in response to the insertion of a 5-fold (red) or 7-fold (blue) disclination. (b) The differing geometries can be characterized by comparing the radial distance between two disks $R$ (dashed black line), and the circumferential distance from an outer disk to itself $C$ (solid purple line). For a flat surface $C = 2 \pi R$ (center), but for a positively curved surface $C < 2 \pi R$ (left), while a for a negatively curved surface $C > 2 \pi R$ (right). (c) Equivalent geometries exist for flexible columnar materials when we consider the distance of separation perpendicular to the filaments. A pattern of twist (left) reduces the circumferential distance between filaments (solid purple), similar to a dome. Alternatively, a pattern of splay (right) reduces the radial (dashed black) distance between filaments, similar to a saddle. (d) Based on these geometries, conjectured structures are shown for a bundle with a 5-fold disclination (left), and 7-fold disclination (right).}
\label{fig:FirstLook}
\end{figure}

We exploit this analogy between 2D crystals and columnar materials to characterize the structural instabilities triggered by the stress from topological defects in the cross-sectional order of bundles. Specifically, we consider the instabilities driven by elementary 5-fold (positive) and 7-fold (negative) disclinations in otherwise hexagonally ordered bundles, characterized respectively, by the removal or insertion of a $60^\circ$ wedge of crystalline material \cite{Chaikin2000}. Just as in 2D crystals, disclinations and dislocations, can be characterized via a Volterra construction corresponding to the mismatch of lattice rotation and displacement around a closed loop encircling the defect.  Disclinations are quantified by a {\it topological charge}, $s$, measuring the angular turning of lattice directions around the defect, which must be integer multiples of $2 \pi/6$ so that the array remains 6-fold at all points except defect cores. The analogous problem, the shape of hexagonally ordered membranes with 5- and 7-fold disclinations, was studied by Seung and Nelson, in the context of the F\"{o}ppl-von K\'{a}rm\'{a}n (FvK) theory of crystalline sheets \cite{Seung1988a}. Creating a 5-fold (7-fold) disclination requires removing (adding) a wedge of crystalline material from the sheet, which stretches (compresses) distances azimuthally around the defect, yielding tensile (compressive) stresses along the hoop direction. These stresses are mechanically balanced by compressive (tensile) stresses along the radial lines extending from the disclination. The fact that thin elastic sheets are unstable to compression now justifies the shapes favored by disclinations: Five-fold defects favor conical shapes---with positive Gaussian curvature---buckled along the radial lines extending from the defect; while 7-fold defects favor saddle shapes---with negative Gaussian curvature---buckled along the azimuthal hoops surrounding the defect (shown visually in Fig.~\ref{fig:FirstLook}(a)). The buckling behavior of a crystalline sheet of radius $R$ is governed by a single dimensionless number, the F\"{o}ppl-von K\'{a}rm\'{a}n (FvK) number, $\gamma_s \equiv Y R^2/B$, which characterizes the crystal's relative resistance to in-plane stretching vs.\ out-of-plane bending, described by the elastic moduli $Y$ and $B$, respectively. Disclinated crystals remain flat for small $\gamma_s$, but become unstable above a threshold value (which is somewhat higher for 7-fold than 5-fold), at which point in-plane stretching exceed the cost of bending to yield a buckled 3D shape.

Complementary to this mechanical perspective is the geometrical one, in which topological defects redefine the metric of the 2D surface in which the crystal is embedded \cite{Moshe2015}. Consistent with the Gauss-Bonnet theorem \cite{Kamien2002a}, the deficit or excess angle associated with disclinations can be accommodated without far-field strain, provided that it is balanced by the integrated Gaussian curvature of the sheet. This is illustrated in Fig.~\ref{fig:FirstLook}(b), where the relative lengths of the radial (dashed black) and circumferential (solid purple) paths along the surface depends on the curvature. Similarly, in a columnar material, variations in filament orientation can be linked to geometrical constraints on their spacing \cite{grason2010a, grason2015a}, which in turn has a precise connection to the metric geometry. This can be seen in the simplified depictions of filament bundles in Fig.~\ref{fig:FirstLook}(c), which also shows the radial (dashed black) and circumferential (solid purple) distances between filaments. The result is a unique equivalence between a pattern of filament orientation and a corresponding surface with a Gaussian curvature of
\begin{equation}
K_{\rm eff} \simeq \frac{1}{2} \left[ \partial_x^2 (t_y)^2 + \partial_y^2 (t_x)^2 - 2 \partial_x\partial_y (t_x t_y) \right],
\label{eq:KGEquivCont}
\end{equation}
where $t_x$ and $t_y$ are respectively the $x$ and $y$ components of filament tilt in the plane normal to their mean orientation, $\hat{z}$.

Given the generic instability of columnar structures to the internal tensile stresses generated by disclinations, it is reasonable to expect that sufficiently flexible bundles will buckle into non-parallel 3D shapes. However, it is {\it a priori} unclear exactly what 3D patterns of orientation will be trigger by such defect-generated stresses, nor what parameters control their relaxed shapes (as $\gamma_s$ does for crystalline sheets). Beyond their mechanical and geometrical correspondence as 2D ordered materials, crystalline membranes and columnar bundles have obvious and profound differences. Specifically, bundles are fully three-dimensional structures, i.e.~their full degrees of freedom are not reducible to a 2D manifold. And it remains to be understood how their buckling behavior relates to their well-studied 2D membrane counterparts. For example, assuming the pattern of tilt to be axisymmetric in response to a centered disclinations, one might anticipate that 5- and 7-fold defects generate the respective 3D double-twist (left) and undulating splay (right) tilt patterns shown in Fig.~\ref{fig:FirstLook}(d) \cite{grason2015a}. As we describe below, the distinction between such shape modes gives rise to vastly different elastic energies. This reveals a fundamental asymmetry between the ability of bundles to realize the analogs of {\it positive-} or {\it negative-curvature} metric geometries; and as a consequence, we observe a profoundly asymmetric response to these two elementary defect types.

In this paper, we employ a combination of continuum elastic theory and discrete-filament simulations of a minimal model of cohesive bundles to study the shape transitions driven by single elementary 5- or 7-fold disclinations. Based on an axisymmetric model of infinite-length bundles, we show that the buckling behavior of bundles is governed by a parameter that we a call the {\it filamentary}-von Karman ({\it fil}-vK) number. Analogous to the FvK number for membranes, the {\it fil}-vK measures the dimensionless ratio of inter-columnar distortions (imposed by defects) to the cost of the lengthwise bending of filaments. However, unlike the case of crystalline sheets, we show that 5-fold defects lead to shape buckling of {\it all} bundles, without a threshold for size or filament flexibility. In contrast, we show that 7-fold defects are characterized by a finite {\it fil}-vK number instability threshold for axisymmetric splay undulations (i.e.\ finite bundle diameter or filament stiffness). We show that this dramatic asymmetry derives from the existence of a uniquely soft torsional mode available to 5-fold defects that generates an equivalent positive curvature without lengthwise variation in strains. Whereas for 7-fold defects, no such shape mode exists that provides ``negative-curvature'' without breaking longitudinal (i.e.\ lengthwise) symmetry, and thus amplifying inter-column strains.  Indeed, we find that this additional frustration of negative-curvature underlies a much more profound symmetry-breaking for 7-fold bundles that can be described with the simplistic assumption of axisymmetry. Comparing our continuum analysis to discrete-filament simulations we find that for 5-fold defective bundles composed of very flexible filaments, there is a spectrum of metastable torsionally-wrinkled modes. While for 7-fold defective bundles, we find significantly more complex non-axisymmetric modes that compromise the drive for ``negative-curvature'' with the preference for uniform inter-filament strains along the length. We argue that these low-symmetry ``counter-twisted'' tilt patterns allow for a significant reduction of the threshold diameter for buckling 7-fold defective bundles, yet a finite threshold must remain in the infinite length limit due to the breaking of lengthwise symmetry. Nevertheless, these results are consistent with the qualitative distinctions captured by the asymmetric shape modes. Finally, we conclude with a discussion of {\it fil}-vK number values and their ramifications for various experimental systems of cohesive filament bundles.

\section{Axisymmetric shape instabilities}
\label{sec: continuum}
In this section we explore a continuum elastic description of a columnar bundle possessing a 5- or 7-fold disclination in its cross section. We then analyze how these defects trigger axisymmetric shape instabilities. While we will show in Sec.\ \ref{sec:ModelIntro}, that the assumption of axisymmetry ultimately fails for 7-fold defective bundles due to frustration between negative-curvature geometry and longitudinal symmetry, the analysis of the axisymmetric shape modes most clearly illustrates the mechanical and geometric principles that underly a profoundly asymmetric response to positive vs.\ negative disclinations. Furthermore, it highlights the critical combination of the material parameters that govern the shape response of bundles to defects. Our analysis examines the case of a hexagonally ordered columnar material, which can be considered to be a generic example of a filament bundle.

\subsection{Continuum elasticity of defective bundles}
Consider an initially cylindrical bundles with a radius $R$ and length $L \to \infty$ along the $\hat{z}$ axis. Here, as in refs. \cite{grason2010a, grason2012c}, the stress-free reference state is described as a 2D hexagonal array of parallel filaments. Deformations are measured relative to the reference configuration by the displacements ${\bf u}_\perp (\xv)$ of local filaments at a point $\xv$ in the bundle. As there is no cost for lengthwise displacements in such a material, ${\bf u}_\perp (\xv)$ is 2D and perpendicular to $\hat{z}$. The elastic energy for deformation is
\begin{equation}
E_{elas} = \frac{1}{2} \int dV (\lambda u^2_{ii} + 2\mu u_{ij} u_{ij}),
\label{eq:ContElEn}
\end{equation}
where $\lambda$ and $\mu$ are the Lam\'{e} coefficients deriving from inter-filament cohesive forces \cite{Selinger1993, Oswald2005}, and related to the 2D Young's modulus, $Y= 4 \mu (\lambda + \mu) / (\lambda + 2 \mu)$, and Poisson ratio, $\nu = \lambda/(\lambda + 2 \mu)$ of the filament array. The strain tensor has components in the $xy$ plane~\footnote{We focus on the limit of small in-plane displacement gradients ($|\partial_i {\bf u}_\perp| \ll 1$), and hence, neglect higher-order contributions to the strain tensor proportional to $\partial_i {\bf u}_\perp \cdot \partial_j {\bf u}_\perp$.}
\begin{equation}
u_{ij} = \frac{1}{2} (\partial_i u_j + \partial_j u_i - t_i t_j),
\label{eq:ContElStrain}
\end{equation}
where $t_i$ is the in-plane component of the filament tangent unit vector. In the limit of small tilt ($|\partial_z {\bf u}_\perp|\ll 1$), the tangent is
\begin{equation}
\label{eq: tangent}
{\bf t} (\xv) \simeq \hat{z} + \partial_z {\bf u}_\perp.
\end{equation}

Additionally, we consider the elastic cost of lengthwise gradients of $\tv (\xv)$ associated with filament curvature $\kappa = |(\tv \cdot \grad) \tv|$~\footnote{The most general model includes terms in the form of the Frank elastic free energy density, $\big[K_1(\grad \cdot \tv)^2 + K_2| \tv \cdot (\grad \times \tv)|^2 + K_3 |(\tv \cdot \grad) \tv|^2\big]/2$, with respective penalties for splay, twist, and bend. Here, we focus on the simplest model where constituent filaments have intrinsic stiffness, but lack strong orientation-dependences of cohesive interactions that would generate ``bare'' costs for splay or twist, beyond the elastic cost for lengthwise in-plane strains generated in columnar materials (typically interpreted as a divergence of the renormalized values of $K_1$ and $K_2$ as $L\to \infty$ in columnar systems).}
\begin{equation}
E_{bend} = \frac{K}{2} \int dV |\partial_z \tv|^2,
\label{eq:ContElBend}
\end{equation}
where the value of the Frank constant, $K$, is proportional to the intrinsic bending modulus of filaments, $B$. The ratio of intra- to inter-filament elastic moduli defines a length scale $\lambda_b^2 \equiv K/Y$, typically associated with the {\it penetration depth} of bending deformation in a columnar material \cite{Jonathan1991}. Comparing this size scale to the lateral size of the bundle defines the dimensionless {\it fil}-vK number,
\begin{equation}
\gamma \equiv \Big(\frac{ R}{\lambda_b} \Big)^2 = \frac{ Y R^2}{K}.
\end{equation}
Analogous to the FvK number for thin membranes $\gamma_s$, $\gamma$ assesses the relative costs of inter-filament vs.\ intra-filament deformations in the bundle, and as we show below, is critical for regulating the buckling of unstable bundles.

To explore the connection between the defect-induced instabilities of crystalline membranes and columnar bundles, we first consider the Euler-Lagrange equations of $E = E_{elas} + E_{bend}$, for the case of a tilt pattern $\tv(\xv)$ that is fixed along the length ($\delta \tv(\xv)=0$). These are simply conditions of in-plane force balance,
\begin{equation}
\label{eq: fixedt}
\Big( \frac{ \delta E}{ \delta u_j (\xv)}\Big)_{\delta \tv(\xv)=0} = - \partial_i \sigma_{ij} =0,
\end{equation}
with a stress of $\sigma_{ij} = \delta_{ij} \lambda u_{ii} + 2 \mu u_{ij}$. Like the case of a membrane with a fixed topography, the relaxation of the displacement $\uv_\perp$ for bundles can be derived from the conditions of in-plane force balance augmented with a compatibility equation. This equation enforces the stress contributions from both the in-plane components of $\tv(\xv)$ as well as singularities in the displacement fields associated with topological defects in the 2D crystalline order,
\begin{equation}
\label{eq: compat}
Y^{-1} \grad_\perp^2 \sigma_{ii} = s(\xv) - K_{\rm eff} , \ \ {\rm for} \ \delta \tv(\xv) = 0,
\end{equation}
where $s(\xv)=\sum_\alpha s_\alpha \delta^{(2)} (\xv-\xv_\alpha)$ is the areal density of a topological disclination charge ($s_\alpha$ is the charge and $\xv_\alpha$ is the position of the disclination $\alpha$), and $K_{\rm eff}$ is the Gaussian curvature of a surface that approximates the inter-columnar metric of the 2D bundle cross section \cite{grason2015a}. Illustrated visually, Fig.~\ref{fig:FirstLook}(b) shows a 2D surface that approximates the inter-filament distances found in Fig.~\ref{fig:FirstLook}(c). Hence, both topological defects ($s(\xv)$) and filament tilt patterns for which $K_{\rm eff}\neq 0$, act as sources for far-field inter-filament stresses. This relation has been previously used to show, for example, that positively charged (5-fold) disclinations become stable for bundles with fixed and sufficiently large double-twist (i.e.~the tilt pattern shown on the left in Fig.~\ref{fig:FirstLook}(d)).

However, there is a complication in determining the buckling modes for negatively charged (7-fold) disclinations: this unusual case of double-twist is the only tilt pattern that yields a constant strain along the bundle's length; but this pattern alone creates the wrong effective curvature (positive rather than negative). Alternatively, while a locally-splayed geometry where $K_{\rm eff} < 0$, may partially neutralized a negatively charged (7-fold) disclinations---the right image in Fig.~\ref{fig:FirstLook}(d)---this tilt pattern unfavorably breaks lengthwise symmetry (i.e.\ because $\partial_z \uv_\perp = \tv_\perp$). In Sec.\ \ref{sec:linStab}, this axisymmetric splay pattern is used to determine the generic $\gamma$-dependence of buckling 7-fold defective bundles. Though, in Sec.\ \ref{sec:ModelIntro}, our discrete filament simulations reveal that frustration between negative curvature and longitudinal symmetry leads to a far more complex and lower energy tilt pattern that preempts the axisymmetric instability for 7-fold, but not 5-fold, defective bundles.  We will show that the breaking of lengthwise symmetry cannot be avoided for 7-fold defective bundle, and thus the qualitative and profound distinctions between 5- and 7-fold defective bundles predicted by the continuum analysis, are still borne out by simulations with unconstrained shapes.

Our purpose is to understand the equilibrium patterns of displacement (and thus orientation) that result from fixed topological defect structure. Therefore, to accurately determine mechanical equilibrium through the Euler-Lagrange conditions, we must consider lengthwise variation of $\uv_\perp$ and the associated variation of $\tv_\perp \simeq \partial_z \uv_\perp$. These more general equilibrium conditions used in the instability analysis below take the form, 
\begin{equation}
\label{eq: EL}
\frac{ \delta E}{ \delta u_j (\xv)} = - \partial_i \sigma_{ij} + \partial_z \big[ t_i \sigma_{ij} \big] +K \partial_z^3 t_i = 0 .
\end{equation}
Relative to the case of a fixed tilt pattern in eqn (\ref{eq: fixedt}), this force balance introduces two additional terms. The first term, $\partial_z \big[ t_i \sigma_{ij} \big]$, couples stresses in consecutive ``layers'' of the bundle, and it is the analog of the ``Young-Laplace" contribution to the normal force (i.e.~in the first FvK equation) from in-plane stresses in curved membranes. The last term, proportional to $K$, derives from torques generated by bending of the filaments, which are expressed here as in-plane forces.

\subsection{Linear stability of parallel, defective bundles}
\label{sec:linStab}
Given this foundation of a continuum elastic model of columnar materials, we now employ linear stability to determine the buckling patterns caused by centered disclinations. The results will reveal a fundamental difference in the charge (7- vs 5-fold) dependence of deformation, owing to symmetry breaking in the lengthwise direction, an aspect unique to columnar materials.

To begin, we consider the stability of an initially parallel bundle ($\tv_0 = \hat{z}$) possessing a centered disclination, whose equilibrium stress $\sigma_{ij}^0$ satisfy eqns (\ref{eq: compat}) and   (\ref{eq: EL}), 
\begin{equation}
\sigma^0_{rr} = \frac{Y s}{4 \pi} \ln(r/R),  \quad \sigma^0_{\phi \phi} = \frac{Y s}{4 \pi} \big[\ln(r/R)+1\big],
\end{equation}
and $\sigma_{r \phi} =0$, where $r$ and $\phi$ are respectively the cross-sectional in-plane radial and polar angle coordinates, and $s=\pm \pi/3$ is the topological disclination charge (where $\pm$ refers respectively to 5- and 7-fold defects). From this base displacement field, $\uv_0$, generated by the defect (associated with $\sigma_{ij}^0$), we apply non-parallel displacements of  $ \delta \uv(\xv)$, such that $\uv(\xv) = \uv_0 (\xv) + \delta \uv(\xv)$. In particular, we consider deformations that are periodic along $z$ and {\it axisymmetric} in the plane,

\begin{equation}
\delta \uv (\xv) = \delta u_r (r) \cos (kz) \hat{r} + \delta u_\phi (r) \cos (kz) \hat{\phi}.
\end{equation}
These two periodic shape modes we refer to as, {\it splay-} and {\it torsional wrinkles} respectively, for $k\neq 0$. We can now analyze the instability of a parallel bundle to splay ($\delta u_r\neq 0$) or torsional ($\delta u_\phi\neq 0$) shape modes derived from the existence of solutions to the force balance equations. Naturally, we consider the limit of small deflections from the initial parallel state; or in other words, the solution to eqn (\ref{eq: EL}), to linear order in $\delta \uv (\xv)$. 

The (linearized) force balance equations can be recast in a simple and surprisingly familiar form (see Appendix \ref{sec:LinearStab}),
\begin{equation}
\Big[-\frac{R^2}{2} \grad_r^2 +V_\alpha(r)\Big] \delta u_\alpha = -\epsilon_\alpha \delta u_\alpha,
\label{eq: coulomb}
\end{equation}
where $\grad_r^2 f= \partial_r \big[ r^{-1} \partial_r( rf) \big]$ is the radial part of the 2D Laplacian, and the exact forms of $V_\alpha(r)$ and $\epsilon_\alpha$ are given in 
eqs. (\ref{eq: potential}) and (\ref{eq: eigen}).  The ``potential", $V_\alpha(r)\propto - s (kR)^2(\ln r + C_0)$, derives from the coupling of tilt to the defect-induced stress, while the eigenvalue, $\epsilon_\alpha \propto (k R)^4/\gamma$, derives from forces induced by bending. Thus, the stability of defective bundles to axisymmetric splay- or torsionally-wrinkled shapes are formally equivalent to finding (zero angular momentum) bound states of a 2D hydrogen atom ``energy" $-\epsilon_\alpha$, whose central ``charge" is $s k^2$. The boundary conditions are determined by the condition of finite stress at the center of the bundle, or $\delta \uv (0) = 0$, and vanishing stresses at the outer bundle surface, which take the form for radial stresses
\begin{equation}
\Big(\partial_r \delta u_r + \nu \delta u_r /r \Big)_{r=R}= 0,
\end{equation}
and azimuthal stresses
\begin{equation}
\Big(\partial_r \delta u_\phi - \delta u_\phi /r \Big)_{r=R}= 0.
\end{equation}
While superficially similar in form, the distinct boundary conditions underly profound difference between splay and torsional deformations of 2D columnar materials. Torsional modes with ``zero kinetic energy", that is $\delta u_\phi \propto r$, generate no shear stress at the bundle surface, while the same is not true corresponding to radial modes, which indicate that splay ground states acquire ``kinetic energy''. As a result, for splay modes the existence of ``bound states", where $\epsilon_\alpha \geq 0$, occurs only for finite $k$ where $V_r(r)$ is sufficiently strong, while for torsional modes bound states exist for all wave vectors down to $k\to 0$. 

\begin{figure}[t]
\centering
\includegraphics[width=0.48\textwidth]{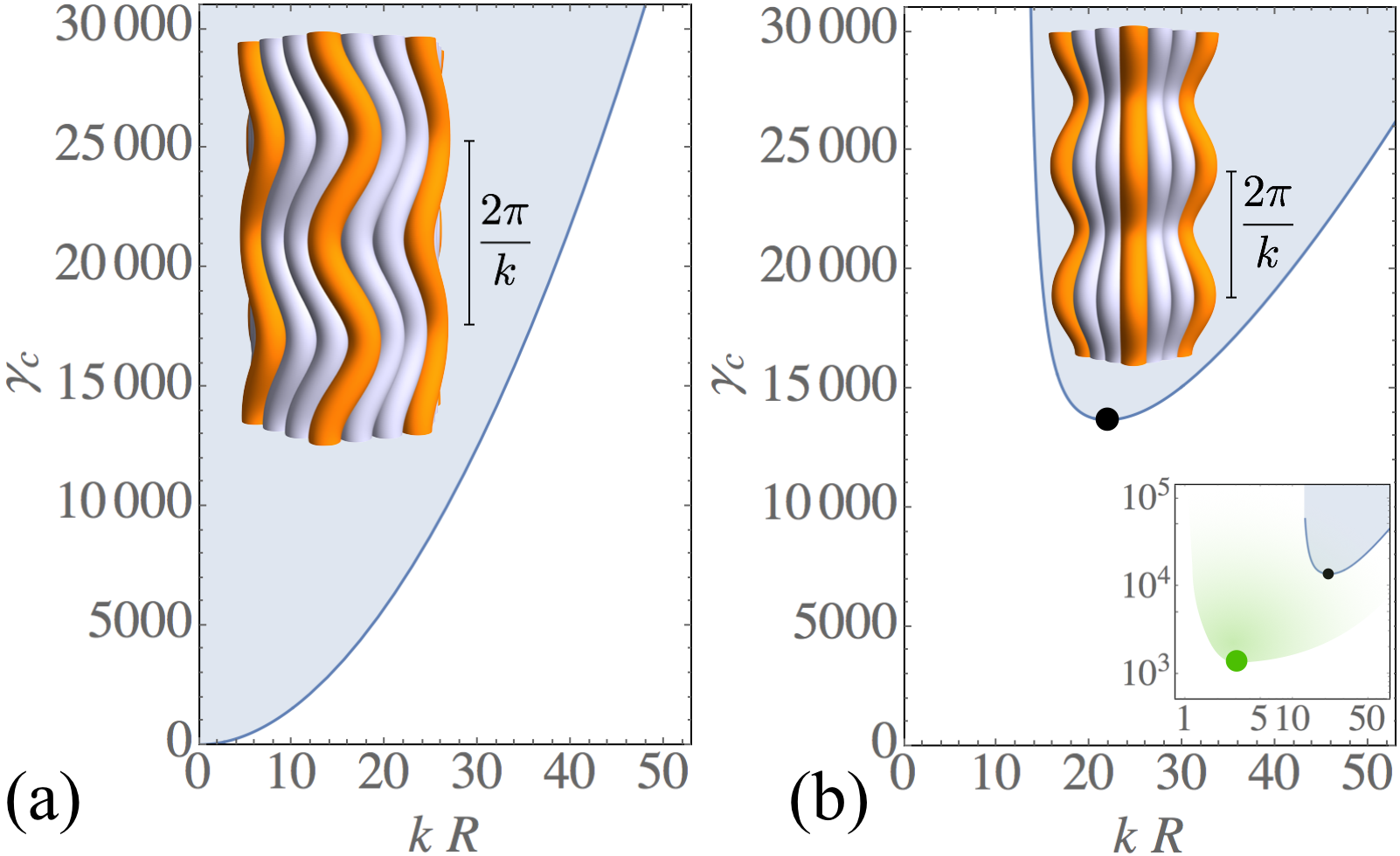}
\caption{The stability map of torsional wrinkles for a bundle in terms of the wave vector, $k$, and the {\it fil}-vK number, $\gamma$. Above this line, bundles are unstable to torsional wrinkling. Visualizations of the unstable modes are shown for each case. (a) A 5-fold disclination yields a stability range that decreases to zero in the long-wavelength limit of $k \rightarrow 0$, i.e.\ homogeneous pitch. (b) Alternatively, a 7-fold disclination is unstable to splay undulations. The minimum $\gamma$ unstable mode (black dot) occurs for $kR = 21.8$, and $\gamma_c = 13,685$. The inset shows a schematic comparison between the axisymmetric instability and the $\gamma_c$ corresponding to the non-axisymmetric shape modes explored by the discrete filament simulations (green dot), analyzed in Sec.\ \ref{sec:7foldSim} and extrapolated to the $L \rightarrow \infty$ limit.}
\label{fig:LinStabPred}
\end{figure}

The results of this ground state analysis is shown in Fig.~\ref{fig:LinStabPred} (for $\nu=1/3$, chosen to make comparisons to our discrete model later on). Here $\gamma_c (k)$ shows the critical {\it fil}-vK number, above which 5- and 7-fold defective bundles are unstable to torsional and splay instabilities at a wave vector $k$. The distinct wave vector dependence of these instabilities follows from simple energetic arguments: Consider first, a torsional mode, $\delta u_\phi \approx \tau_0  r \cos (k z)$, where $\tau_0$ is a constant. Because this mode is a purely rigid rotation around $\hat{z}$, to linear order elastic strains vanish (i.e.~$\partial_i u_j+\partial_ju_i = 0$), and the only elastic strains are generated by tilt $\delta {\bf t} \approx \tau_0 k r \hat{\phi}$, leading to a mean strain $\langle \delta u_{\phi \phi} \rangle \approx - k^2 \tau_0^2 r^2 $ that is compressive in the hoop direction. This leads to a relaxation of defect induced stress of $\int dA ~ \sigma^0_{\phi \phi} \langle \delta u_{\phi \phi} \rangle > Y s k^2 R^4 \tau_0^2$, which is dominated by large $r$ where $\sigma^0_{\phi \phi}(r)/s>0$. Combining this with the bending cost gives an energy density $\varepsilon_{{\rm tor}}$ for torsional-wrinkling (relative to the parallel state)
\begin{equation}
 \Delta \varepsilon_{{\rm tor}} \approx \big [ -Ys k^2 R^ 2+ K k^4 R^2\big] ~ \tau_0^2 + {\cal O} ( \tau_0^4).
\end{equation}
This shows that the relaxation of tensile strain generated by 5-fold defects ($s=+\pi/3$) exceeds the bending cost for modes, $k < k_c \approx \lambda_b^{-1}$, or $\gamma_c \sim (k R)^2$, with the scaling shown in Fig.~\ref{fig:LinStabPred}(a). Hence, bundles of any size or stiffness are unstable to long-wavelength ($k \to 0$) torsional-wrinkles. In infinite bundles, such modes correspond to uniform helical twist $\delta {\bf t} \simeq \Omega r \hat{\phi}$, studied previously as an {\it ansatz} for elastic energy ground states in the presence of 5-fold disclinations \cite{grason2010a}.

As $L \to \infty$, this lack of a threshold for the shape instability driven by 5-fold defects is unlike the analogous problem of conical buckling of crystalline sheets. This difference derives from the fact that in the latter case, the elastic energy released by conical buckling is proportional to the square of the sheet curvature (i.e.\ the Gaussian curvature), just as the (positive) energy cost of bending. Hence, membrane stiffness must fall below a critical value for shape buckling. However, for cohesive bundles, the specific shape mode driven by positive disclinations is a {\it soft mode}, generating elastic elastic costs only at ${\cal O}(\Omega^4)$, which for small twists, are always overwhelmed by the elastic energy released by twist, proportional to $s \Omega^2$.

Turning to the case of 7-fold defects ($s = - \pi/3$), the energetic analysis proceeds along similar lines for a shape mode ansatz of radial splay, which we approximate by the linear profile $\delta u_r \approx \rho_0 r \cos (k z)$, where $\rho_0$ is a constant. However, unlike torsional wrinkles, splay leads not only to tilt-induced radial strains ($\propto (k \rho_0 r)^2$), but also to linear in-plane strains, due to area dilation $\grad_\perp \cdot \delta {\bf u} \sim \rho_0 \cos (k z)$, generating an elastic cost (per unit volume) of $\approx Y \rho_0^2$. Thus, the cost of splay modes has the form,
 \begin{equation}
\Delta \varepsilon_{{\rm splay}} \approx \big[ Y + Y s k^2 R^2+K k^4 R^2\big] \rho_0^2 + {\cal O} ( \rho_0^4),
\end{equation}
where the cost at $k \to 0$ derives from the well-known coupling between splay and lengthwise density variations in columnar systems. The radial splay modes that relax elastic energy (due to the collapse of the radial tension generated by 7-fold defects) are unlike twist in that they are not soft modes. The balance between relaxing bending and elastic energy selects an optimal wrinkling wavelength $k_c \approx |s|^{1/2} \lambda_b^{-1}$, with a net relaxation proportional to $-Ys^2 \gamma$. Thus, only when the {\it fil}-vK number exceeds a threshold value, will the 7-fold defect drive (finite $k$) splay-wrinkling of the bundle. This result for 7-fold defects is shown in Fig.~\ref{fig:LinStabPred}(b). Unlike the torsional wrinkling in the presence of a 5-fold defect, which becomes unstable at long-wavelengths, now there is a range of long-wavelengths (small $k$), for which no unstable solution exists at any $\gamma$. The minimum unstable value of $\gamma_c$, occurs at a mode $kR = 21.8$, for which $\gamma_c = 13,685$, setting an upper limit threshold {\it fil}-vK number for the splay instability driven by a 7-fold disclination. For large $k$, the stability line again follows the scaling of $\gamma_c \sim (k R)^2$.  

The stability analysis to axisymmetric shape modes illustrates a profound asymmetry between the response to 5- vs.\ 7-fold defects in the infinite length limit. Bundles with centered 5-fold defects are always unstable; while for 7-fold defects, parallel bundles are stable up to a finite $\gamma$ (proportional to their lateral area), beyond which they become unstable to lengthwise shape modes at a finite wavelength. This is in sharp contrast to crystalline membranes, where the modest asymmetry in the bending cost of respectively conical vs.\ saddle-like shapes driven by 5- and 7-fold defects lead to only slight difference in the critical FvK number for buckling. For bundles, this dramatic asymmetry can be attributed to the fact that ``double-twist" generates the equivalent of positive curvature geometries ($K_{\rm eff}>0$), but requires no strain variation along the bundle's length. In fact, it can be shown that the uniform double-twist pattern is the {\it only} texture that does not break lengthwise symmetry~\cite{Atkinson2017}. Therefore it is the only soft mode available for deformation (i.e.\ available to any $\gamma \neq 0$) because it doesn't lead to the immense strains caused by area dilation. Consequently, generating negative equivalent curvature ($K_{\rm eff}<0$) through axisymmetric splay leads to an elastic cost for such modes that does not vanish in the $k\to 0$ limit.

In Sec.\ \ref{sec:7foldSim} we find that 7-fold defective bundles are still unstable to modes that (must) break lengthwise symmetry; however there exists a more exotic tilt pattern of lower energy that allows for buckling at a lower (though necessarily still nonzero) $\gamma$. As a consequence, the asymmetric splay mode that was described analytically in the current section is in fact preempted by this non-axisymmetric shape mode, shown schematically in the inset to Fig.~\ref{fig:LinStabPred}(b). Although this instability is triggered earlier ($\gamma_c \approx 1,500$ and $kR \approx 3$), we will show below that the negative-curvature tilt pattern breaks lengthwise symmetry as required, and therefore imposes a finite value for the critical {\it fil}-vK number. This behavior is in stark contrast to 5-fold defective bundles that are unstable for all $\gamma$.

As alluded to above, the linear stability analysis of axisymmetric modes does not necessarily capture the true symmetries of the most stable deformation pattern, nor the far-from threshold buckled configuration. Additionally, we have ignored the effects of bundle ends by taking the $L/R \rightarrow \infty$ limit. In the following sections, we lift these constraints and compare our analytic results to those from simulations of a finite-length discrete-filament model of cohesive bundles.

\section{Discrete model of cohesive filament bundles}
\label{sec:ModelIntro}
Here we introduce a bead-spring model of cohesive filaments. Our purpose is to determine the elastic energy ground states of bundles containing disclinations, without constraining the symmetry of their deformed shapes, as was done in section \ref{sec: continuum}. This model treats individual filaments as semi-flexible and cohesive ``featureless'' tubes, that incur no elastic cost for lengthwise sliding of neighboring filaments, but do generate costs for lateral deformations that strain inter-filament distances.
\begin{figure}[t]
\centering
\includegraphics[width=0.35\textwidth]{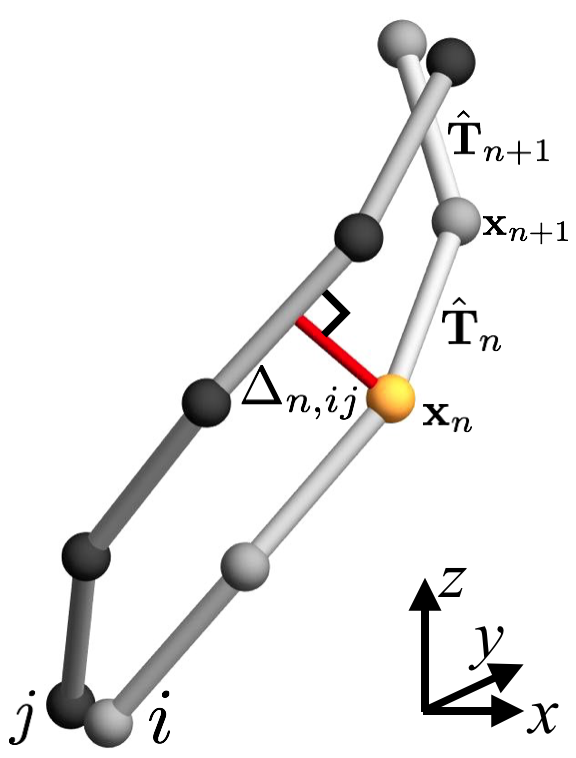}
\caption{Discrete model of cohesive filaments and their interactions, for eqns (\ref{eq:CoarseBend}) and (\ref{eq:CoarseCohesive}). The distance of closest contact (red line) represents the true separation between filaments $i$ and $j$.}
\label{fig:discreteModel}
\end{figure}

A bundle contains $N_{f}$ filaments, indexed by $i=1\ldots N_{f}$, with each filament discretized into $N_{v}$ vertices, or ``beads", indexed by $n=1\ldots N_{v}$. Vertex positions along a single filament are located at the position, ${\bf x}_{i,n}$, and we define $\ell_{i,n}$ as the length of the line segment between vertices $n$ and $n+1$ on filament $i$. The local tangent at $n$ is defined as $\hat{{\bf T}}_{i,n} = ({\bf x}_{i,n +1} - {\bf x}_{i,n}) / \ell_{i,n}$, from which the cost of intrafilament bending is defined
\begin{equation}
E_{b}^{(i)} = B \sum_{n=1}^{N_v-1} \frac{1 - \hat{{\bf T}}_{i,n} \cdot \hat{{\bf T}}_{i,n+1}}{\ell_{i,n}}.
\label{eq:CoarseBend}
\end{equation}
In the limit that $N_v \to \infty$, this energy asymptotically approaches the standard elastic energy for a semi-flexible, worm-like chain.

The elastic cost of cohesive interactions between neighboring filaments $i$ and $j$ are modeled as generic Hookean springs,
\begin{equation}
E_{elas}^{ (i, j)} = \frac{\epsilon}{2} \sum_{n=1}^{N_v} (\Delta_{n, i j } - \Delta_0)^2,
\label{eq:CoarseCohesive}
\end{equation}
where $\Delta_{n, ij}$ represents the {\em distance of closest contact} from vertex $n$ on filament $i$, to a point along filament $j$. This distance intersects $j$ at a right angle as shown in Fig.~\ref{fig:discreteModel}~\footnote{There are rare cases where $\Delta_{n, ij}$ intersects $j$ at a vertex and not a segment, therefore, though not always strictly at a right angle to $j$, $\Delta_{n, ij}$ is always the distance of closest approach.}. In our discretized model, filaments are composed of line segments anchored to jointed vertices, where $\Delta_{n, \langle i j \rangle}$ is calculated between a vertex and its neighboring segment, rather than between vertices~\footnote{At filament ends $\Delta_{n, ij}$ may not perpendicularly intersect filament $j$, thus $\Delta_{n, ij}$ is chosen to point to the end of $j$. This results in end stresses that resist filament orientations not parallel to $z$, but which become negligible for long bundles.}. For sufficiently large $N_v$ this model allows for frictionless sliding between neighbor filaments (particularly when they are straight).

Assembling eqns (\ref{eq:CoarseBend}) and (\ref{eq:CoarseCohesive}), the total free energy of our discrete filament model is
\begin{equation}
E = \sum_{i = 1}^{N_{f}} \Big[ E_{b}^{(i)} + \frac{1}{2} \sum_{\langle i j \rangle} E_{elas}^{(i,j)} \Big],
\label{eq:TotalCoarseEnergy}
\end{equation}
where the final sum is over all the nearest neighbor filaments $j$, to filament $i$. In Appendix \ref{sec:ContinuumVsDiscrete}, it is shown that in the limit of $N_v \to \infty$ and $N_f \to \infty$, the elasticity of this model approaches the continuum limit described by eqns (\ref{eq:ContElEn})and (\ref{eq:ContElBend}) with $\mu=\lambda = \sqrt{3} \epsilon / 2 \ell_0$, $\nu = 1/3$, and $K = 2 B/\sqrt{3} \Delta^2_0$, where $\ell_0$ is the initial intra-filament vertex spacing~\footnote{Note that this discrete filament model does not explicitly generate the equivalents of the Frank elastic costs for local splay and twist (which are also ignored in the continuum model). However, such deformations in general do require strains of the inter-filament Hookean springs, and thus are still penalized.}.

It is straightforward to including an intra-filament stretching cost that favors a constant $\ell_{i,n} = \ell_0$ and acts to maintain inextensibility of filaments. This would be necessary, for example, for a physically accurate description of the interplay between 3D shape and the loss of cohesive contact at the ends of fixed length filaments. However, the present goal is to explore and analyze the buckling behavior of defective bundles, extrapolating to the $L\to \infty$ limit where the effects of boundary interactions are presumably negligible. Hence, filament lengths are not fixed, and rather the vertical $z$ coordinates of vertices are fixed at equally spaced layers (with a separation of $\ell_0$). This has the effect of achieving smaller inter-filament elastic stretching at bundle ends than would occur in the fixed-length case. Strictly speaking, in this model the volume of the bundle and the length of the filaments are no longer conserved. However, this situation is arguably more relevant of certain biological or supramolecular fibers that self-assemble by adjusting length and radius simultaneously. Regardless, we discuss how the non-generic treatment of the bundle ends influences the buckling behavior for finite bundle length.

\subsection{5-Fold Disclinations}
\begin{figure*}[t]
\centering
\includegraphics[width=0.98\textwidth]{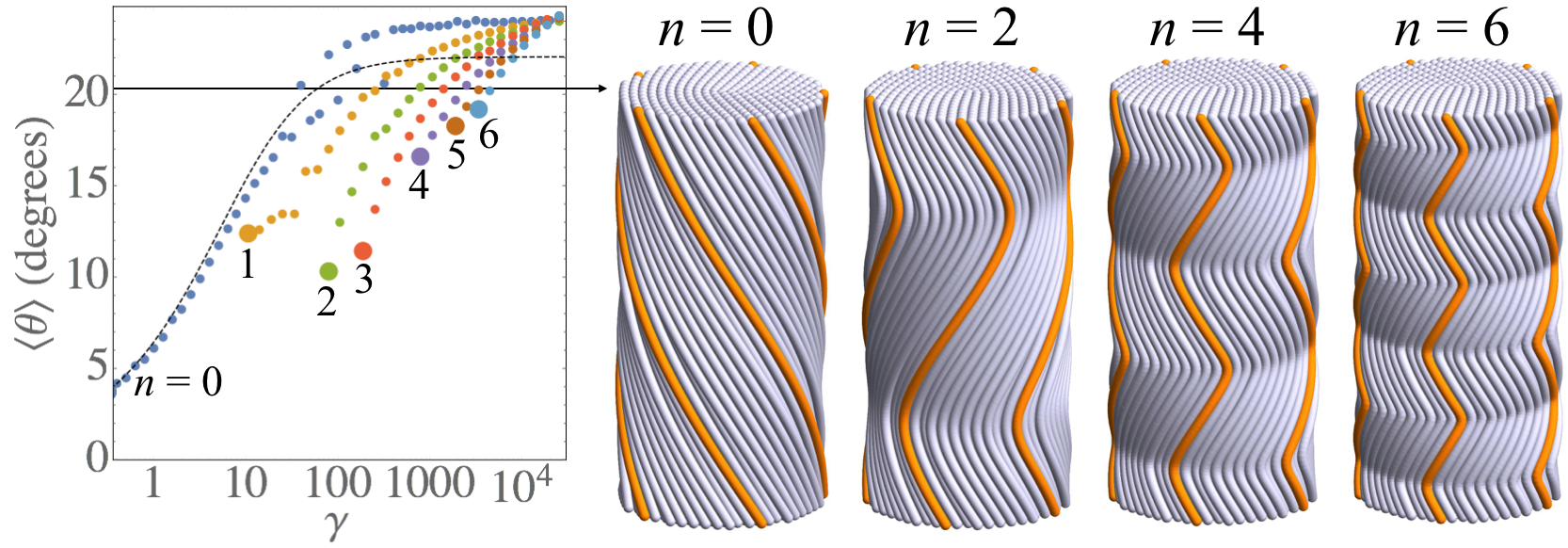}
\caption{Minimal energy results for the discrete model of bundles containing a single centered 5-fold disclination, for various numbers of torsional wrinkles, $n$. The mean twist angle, $\langle \theta \rangle$, is the mean angle of the filaments with respect to the $z$ axis. Dashed black line shows eqn (\ref{eq:GammaEstFit}) for the $n=0$ bundles, with $\gamma_c = 0$, $\theta_0 = 22.1^\circ$, and $\zeta = 32/3$, consistent with continuum elastic theory \cite{grason2010a}. Renderings are shown of twisted bundles for four values of $n$ at $\langle \theta \rangle \approx 20^\circ$. Select outer filaments are highlighted orange for viewing purposes.}
\label{fig:5folddisctwist}
\end{figure*}

Here, we consider the shape transition of bundles with a central 5-fold disclination, introduced through a fixed topology of inter-filament elastic bonds. We focus on the case of a high-aspect ratio with $L/R =8$, as we find that finite-length end effects play a relatively small role in the their buckling behavior. We consider bundles with with $N_f = 306$ filaments (radius $R \approx 10 \Delta_0$) and a vertex spacing of $\ell_0 = 0.2 \Delta_0$. Energy minimization was performed using the GSL conjugate gradient package for C \cite{Galassi2009}. Below we explore the structure and energetics of stable and meta-stable states beginning with highly flexible filaments with $\gamma = 25,000$. Then, after energy minimization, $\gamma$ is reduced by increasing $B$, and the energy is minimized again. This process is repeated over 50 steps (in even logarithmic decrements of $\gamma$) until $\gamma = 0.25$.

\begin{figure*}[t]
\centering
\includegraphics[width=0.98\textwidth]{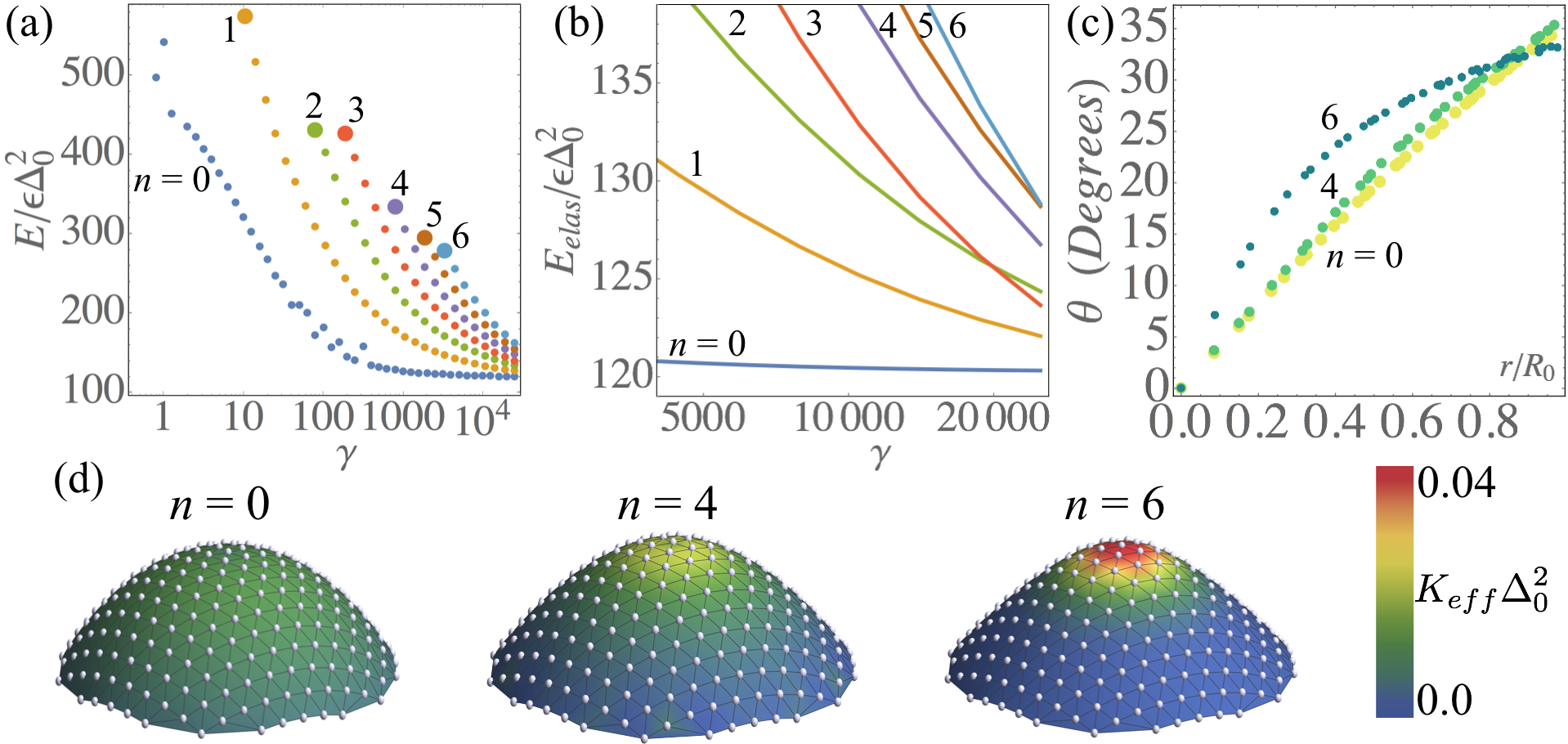}
\caption{(a) Total energy vs {\it fil}-vK number for various $n$. (b) Just the elastic contribution of the total energy, showing a trend for higher $n$ modes overtaking lower ones. For $\gamma = 25,000$: (c) Twist angles of filaments at a distance $r$ from the bundle's center. (d) Reconstructed surfaces (using the method detailed in Appendix \ref{sec:EquivKG}) for select values of $n$, showing how curvature becomes focused near the central defect for higher $n$.}
\label{fig:5foldcurvconc}
\end{figure*}

Results for 5-fold disclinations are shown in Fig.~\ref{fig:5folddisctwist}, where we plot the mean filament twist angle, $\langle\theta\rangle$, defined as the mean angle of all filaments with the centerline (i.e.\ $\cos \theta_{i,n} = \hat{{\bf T}}_{i,n} \cdot \hat{z}$) vs.\ $\gamma$. Data point colors represent different initial configurations generated by applying an azimuthal displacement pattern to all filament vertices of the form $0.2 r \cos (n \pi z / L)$, where $n$ is an integer that counts the number of times the handedness of the helical twist changes along the bundle length. This procedure allows us to bias the lengthwise symmetry of distinct equilibrium shapes. The case of $n = 0$ produces a homogeneously twisted bundle, where all filaments possess an identical pitch, which is the lowest energy state for all tested values of $\gamma$. Note that bundles with 5-fold disclinations are always unstable to homogeneous twist ($n=0$) for all values of $\gamma$, consistent with the $L\to \infty$ linear-stability results described above.

\begin{figure}[b]
\centering
\includegraphics[width=0.48\textwidth]{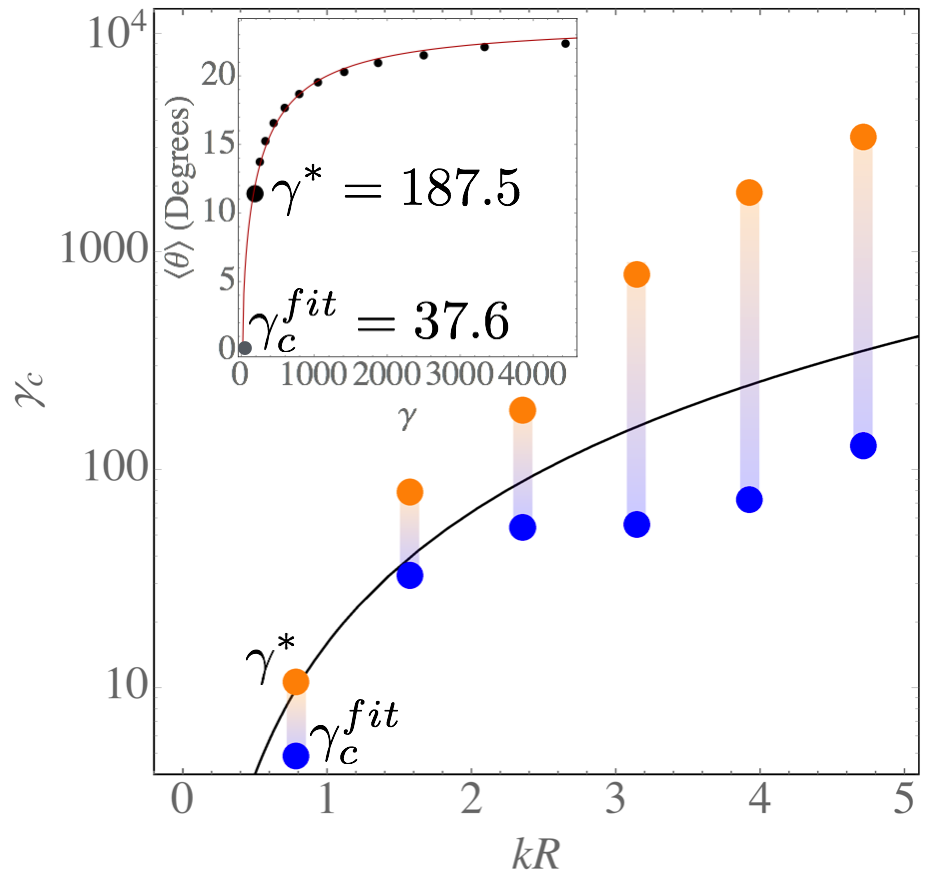}
\caption{Best fit line from eqn (\ref{eq:GammaEstFit}) for the $n = 3$ data from Fig.~\ref{fig:5folddisctwist}. Inset: Confirmation of the linear stability prediction of $\gamma_c$ from Fig.~\ref{fig:LinStabPred}, compared to the upper ($\gamma^\star$) and lower ($\gamma_c^{fit}$) bound estimates for all values of $n = 0$ to 6.}
\label{fig:contDiscComp}
\end{figure}

In addition to uniform pitch states, we see in Fig.~\ref{fig:5folddisctwist} that metastable oscillating twist states are mechanically stable for sufficiently large values of $\gamma$. These torsionally wrinkled shapes are characterized by an alternating direction of twist along the $z$ axis. Our discrete-filament simulations find that for a given $n$-wrinkled shape, there is a value of {\it fil}-vK, designated $\gamma^*(n)$, below which, the bundle becomes unstable to a lower-$n$ structure. These points are highlighted as the large dots in Fig.~\ref{fig:5folddisctwist}(a). As $\gamma$ is decreased, eventually $\gamma^*(n)$ is reached, and the bundle becomes unstable and undergoes a large transition to a new lower energy and lower $n$ of torsional wrinkles. To highlight this trend, Fig.~\ref{fig:5foldcurvconc}(a) shows the total energy vs.\ $\gamma$. For large $\gamma$, the transitions between alternating handedness of twist become sharp kink-like boundaries, consistent with bending being concentrated over length scales proportional to $\lambda_b = R/\gamma^{-1/2}$. 

The region of stability of $n$-wrinkled bundles is consistent with the linear-stability analysis of the continuum model, by assuming that torsional oscillations are commensurate with the finite length of the bundle, or $k_n = \pi n/L$. On one hand, $\gamma_c(k)$, predicted by the continuum model, defines the point at which the straight bundle becomes unstable to torsional wrinkling at wave vector $k$, while $\gamma^*(n)$, measured from simulation results, is the smallest value that an $n$-wrinkled bundle is observed to be stable. These two thresholds always satisfy $\gamma^*(n) > \gamma_c(k_n)$. It is not clear what limits the ability to resolve mechanical equilibrium of the $n$-wrinkled state all the way down to the parallel state (i.e.~$\langle \theta \rangle =0$); presumably this derives from the combination of the inherent precision limit of our discrete-filament model and the vanishing of energetic barriers between nearly unstable and stable modes (with lower $n$). Notwithstanding the loss of stability as $\gamma$ approaches the limit of stability for the $n$-wrinkled mode, we estimate the value of this threshold by fitting the $\gamma$-dependence of an $n$-wrinkled mode to the region near $\gamma \gtrsim \gamma^*(n)$
\begin{equation}
\langle \theta \rangle \approx \theta_0 \sqrt{1+ \zeta/(\gamma - \gamma_c)}^{~-1},
\label{eq:GammaEstFit}
\end{equation}
where $\theta_0$ is the maximum twist angle far from the transition point ($22.1^\circ$ for a 5-fold disclination), and $\zeta$ is a value that regulates the speed of the transition. This particular form has two motivations: first, the expectation that near to the stability threshold $\langle \theta \rangle \sim |\gamma- \gamma_c|^{-1/2}$, a characteristic of a supercritical bifurcation; and second, the predicted $\gamma$-dependence for equilibrium uniform $n=0$ twist of 5-fold defective bundles is expected to have the form of eqn (\ref{eq:GammaEstFit}), with $\theta_0 = 2 \sqrt{3}/9~\mathrm{radians}$, $\zeta = 32/3$ and $\gamma_c =0$ \cite{grason2010a}. This fit is shown to agree well with the $n=0$ results in Fig.~\ref{fig:5folddisctwist} (dashed line). The values of the $\gamma_c \neq 0$ for $n \geq 1$ extracted from fits to eqn (\ref{eq:GammaEstFit}) are shown in Fig.~\ref{fig:contDiscComp}, with the upper and lower bounding estimates of $\gamma^*$ and $\gamma_c^{fit}$, showing reasonable agreement between the predicted dependence $\gamma_c$ on $k$ from the continuum theory.

While the total energy of a wrinkled bundles is always found to be increasing with $n$ for a given $\gamma$, we find evidence that, in the large-$\gamma$ limit, oscillating twist structures for high-$n$ tend towards a lower elastic energy than lower-$n$ structures. Fig.~\ref{fig:5foldcurvconc}(b) shows the elastic contribution to the total energy for large values of the {\it fil}-vK number, which shows that $E_{elas}$ from relatively high-$n$ states (e.g.~$n=6$) tends to decrease faster with $\gamma$ than lower-$n$ structures. Extrapolating this to even larger values of $\gamma$ suggests that for sufficiently flexible filaments, highly-wrinkled bundles ($n \to \infty$) could become the lowest-energy shape equilibria (even lower than the uniform twist state) in the $\gamma \to \infty$ limit, where the cost of filament bending is negligible in comparison to the elastic cost.

Recalling the analogous case of crystalline membranes, Seung and Nelson argued that in the asymptoticly flexible limit of $\gamma_s \to \infty$, the far-field elastic stress of a 5-fold disclination can be completely screened. This stress focusing is achieved by a nearly isometric conical shape that concentrates Gaussian curvature to the disclination position \cite{Seung1988a}. For the torsionally wrinkled bundles, the tilt pattern that concentrates $K_{\rm eff}$ to the bundle center is not the uniformly twisted one, but one where filaments tilt rapidly from the $\hat{z}$ direction within the core of the bundle, and adopt a constant $\theta$ in the outer bulk of the cross section. Fig.~\ref{fig:5foldcurvconc}(c) shows the twist angle $\theta$ as a function of a filament's radial distance from the centerline. While this tilt pattern is possible within a given cross section of the bundle, it requires deviations from the constant helical pitch and introduces shear deformations that grow along the bundle length. Nevertheless we observe that for large $\gamma$, bundle shapes tend towards a similar ``curvature focusing" geometry, made possible the torsional wrinkles. Fig.~\ref{fig:5folddisctwist}(d), shows triangulated surfaces with inter-vertex distances equal to the inter-filament distances in the discrete filament model mid-way between the alternating wrinkles. These surfaces show the progressive focusing of curvature towards the central defect as the number of wrinkles increases with $\gamma$, similar to crystalline membranes (see Appendix \ref{sec:EquivKG} for how to calculate the Gaussian curvature). If true, such a feature would be important for stabilizing nontrivial topologies in crystalline columnar materials.

\subsection{7-fold disclinations}
\label{sec:7foldSim}
\begin{figure}[h!]
\centering
\includegraphics[width=0.47\textwidth]{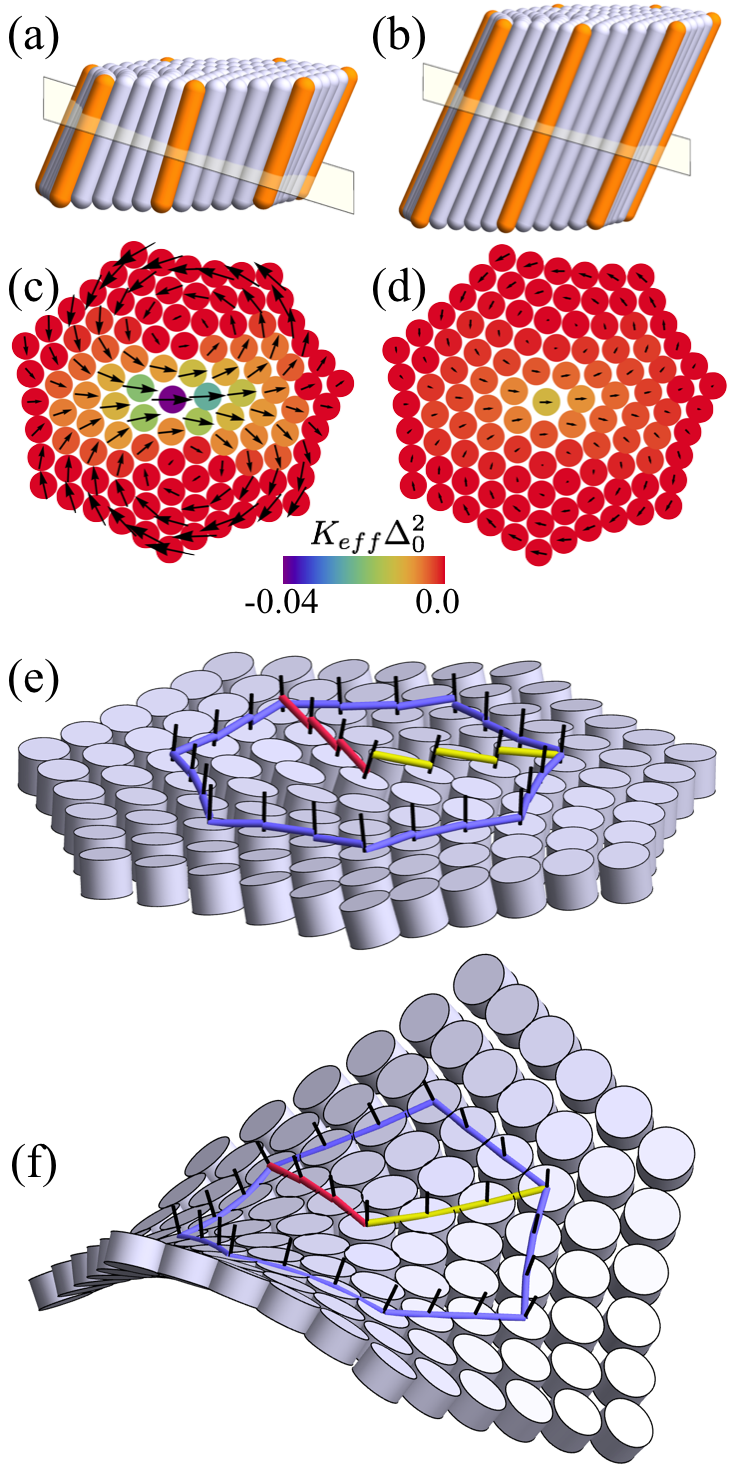}
\caption{Visualizations of infinitely rigid ($\gamma = 0$) filaments with $R/\Delta_0=5$ and (a) $L/R=1$, and (b) $L/R=2$. (c) and (d) show cross-sectional cuts through the bundles, with filaments colored by their local equivalent Gaussian curvature from eqn (\ref{eq:GaussForSurf}). The arrows point in the direction of tilt of each filament away from the average filament orientation. (e) Cross-sectional slice of $L/R=0.4$ bundle, showing a negative Gaussian curvature-like tilt pattern. The colored lines represent the distance of closest contact between neighboring filaments. The blue circumferential and red radial paths are relatively unaffected by tilt, while the yellow radial path is shortened due to filaments tilting. (f) Filaments mapped to an equivalent surface with identical colored paths, but with distances now shortened by curvature rather than filament tilt.}
\label{fig:7FoldRodsImage}
\end{figure}
In this section, we analyze the shape-transitions of bundles with centered 7-fold disclinations. Unlike the torsional wrinkles of 5-fold defects, our discrete-filament simulations show that 7-fold defects lead to buckled shapes that significantly break axisymmetry, in a manner unlike the splay-undulation {\it ansatz} analyzed in Section \ref{sec:linStab}. Nevertheless, despite the different optimal tilt pattern, discrete-filament simulations do reveal that 7-fold defects favor shapes that break the lengthwise symmetry of inter-filament strains, and, as required, only above a critical $\gamma$. However, the tilt pattern is such that area dilation, and therefore costly in-plane strains, are minimized.

Given the complexity of the optimal buckled shapes, we first focus on the limit of infinitely rigid filaments ($\gamma \to 0$) with a finite length, $L$. Bundles of $N_f = 106$ filaments ($R \approx 5 \Delta_0$), of different lengths are shown in Fig.~\ref{fig:7FoldRodsImage}(a-b), with $L = R$ and $L = 2R$ respectively, showing a deformation pattern with two generic features. Most obvious is the tilt of the bundle's centerline with respect to the $z$ axis. Superimposed on this near-uniform tilt is a more subtle pattern of tilt-variation within the bundle's cross section. This pattern is more easily illustrated via the projections of filament tilt in a plane perpendicular to the bundle centerline, as shown in Fig.~\ref{fig:7FoldRodsImage}(c-d). The ``double vortex" pattern viewed in this perspective reveals the surprising emergence of twist driven by 7-fold defects, far from the radial splay pattern assumed on the grounds of axisymmetry. This pattern, which we call the {\it counter twist} tilt pattern, is composed of two double-twisting domains of opposite handedness.

To show that this tilt pattern effectively screens the defect-induced stresses, we analyze the distribution of equivalent Gaussian curvature, $K_{\rm eff}$, using the discrete-filament analysis from Appendix \ref{sec:EquivKG}. Fig.~\ref{fig:7FoldRodsImage}(c-d) shows that regions of $K_{\rm eff}<0$ are predominantly focused at the central 7-fold disclination. One can understand how this counter twist tilt pattern is the metric equivalent to negative Gaussian curvature through the schematic Figs.~\ref{fig:7FoldRodsImage}(e)and (f). Here we highlight distinct inter-filament {\em paths} (colored) in the cross section. In this metric analogy, we are interested in the accumulated distance of closest approach between neighbors spanned along the path, rather than the path length in the planar cross-sectional cut. This distance is dependent on, and can only be shortened by, filaments tilting into the direction of their separation. In Fig.~\ref{fig:7FoldRodsImage}(e), the circumferential (blue) path encompasses the 7-fold disclination, and is relatively unperturbed by filament tilt. However, the radial (yellow) path along the mid-line separating the distinct double-twist domains, is effectively shorted by tilting. Hence, this tilt pattern allows the bundle to shorten distances along (certain) radial directions while keeping the circumferential distance the same relative to a planar geometry, a hallmark of negative curvature geometries. An equivalent negatively curved surface, shown in Fig.~\ref{fig:7FoldRodsImage}(f), shows the same filament tangents and distances of closest approach between them, but now lengths are adjusted by out-of-plane surface curvature rather than in-plane filament tilt. Both patterns are effective at relaxing the compressive hoop stresses generated by 7-fold defects, by expanding the circumference while keep radially separated filaments close to their preferred separation distance.

For this rigid filament case, we note that the tilt pattern is highly sensitive to length. For example, in Fig.~\ref{fig:7FoldRodsImage}(c-d) the tilt variation and $K_{\rm eff}$ decreases significantly from the $L=R$ case to $L=2R$. To understand the length dependence, we analyze this pattern using two quantities: the first measures the tilt of the centerline, $\langle \theta \rangle$, or the mean angle of filaments with respect to the $z$ axis; and the second measures the prominence of the double vortex tilt pattern seen in Fig.~\ref{fig:7FoldRodsImage}(c-d), $\langle \delta \theta \rangle$, defined as the mean tilt angle of filaments away from the centerline. From these parameters we can estimate how the 7-fold bundle energy varies with, $\langle \theta \rangle$ and $\langle \delta \theta \rangle$, as well as $L$ and $R$. (For a derivation of the comparison between our continuum elastic and discrete bead-spring models, see Appendix \ref{sec:ContinuumVsDiscrete}.) We begin with a simple {\it ansatz} for counter twist
\begin{equation}
{\bf t}_\perp \simeq \langle \theta \rangle \hat{x} + \delta \theta\Big[ \Big( \frac{1}{4} - \frac{y^2}{R^2}\Big)\hat{x} + \frac{ x y}{R^2} \hat{y} \Big] ,
\label{eq:StickTilt}
\end{equation}
which has two double-twist patterns centered on $x=0$ and $y=\pm R/2$, and a mean orientation of $\langle \theta \rangle \hat{x}$. Assuming that $\langle \theta \rangle \gg \delta \theta$, we may use eqn (\ref{eq:KGEquivCont}) (and Appendix \ref{sec:EquivKG}) to computed the effective negative curvature of the pattern, giving $K_{\rm eff} \approx - 3 \langle \theta \rangle \delta \theta /R^2$. Hence, the negative curvature geometry relaxes the elastic energy over the bulk of the bundle by an amount $\delta E_{relax} \approx - Y V |s| \langle \theta \rangle \delta \theta$, where $Y \approx \epsilon \ell_0^{-1}$. However, generating this tilt pattern requires two additional costs in rigid filament bundles. First, the tilt variation leads to in-plane strains (dilation and shear) that grow along the bundle length as $u_{ij} \approx \delta \theta z/R$. This cost leads to an elastic penalty that grows rapidly with length $\delta E_{elas} \approx Y V (\delta \theta)^2 (L/R)^2$. Finally, the mean tilt of the bundle axis introduces a stretching cost at the ends of the bundle through the tangential ``slip" of neighbor filaments. For small tilts, the length of the ``slipping" regions are $\ell_{slip} \approx \Delta_0 \langle \theta \rangle$~\footnote{For the rigid, $\gamma \to 0$ limit, we consider the continuous limit along the length, where $\ell_0 \to 0$}, over which the inter filament distance is stretched by an amount $\Delta - \Delta_0 \approx \Delta _0 \langle \theta \rangle^2$, leading to an elastic cost for tilt at the bundle ends of $\delta E_{ends} \approx Y R^2 \Delta_0 \langle \theta \rangle^5$.

\begin{figure}[t!]
\centering
\includegraphics[width=0.48\textwidth]{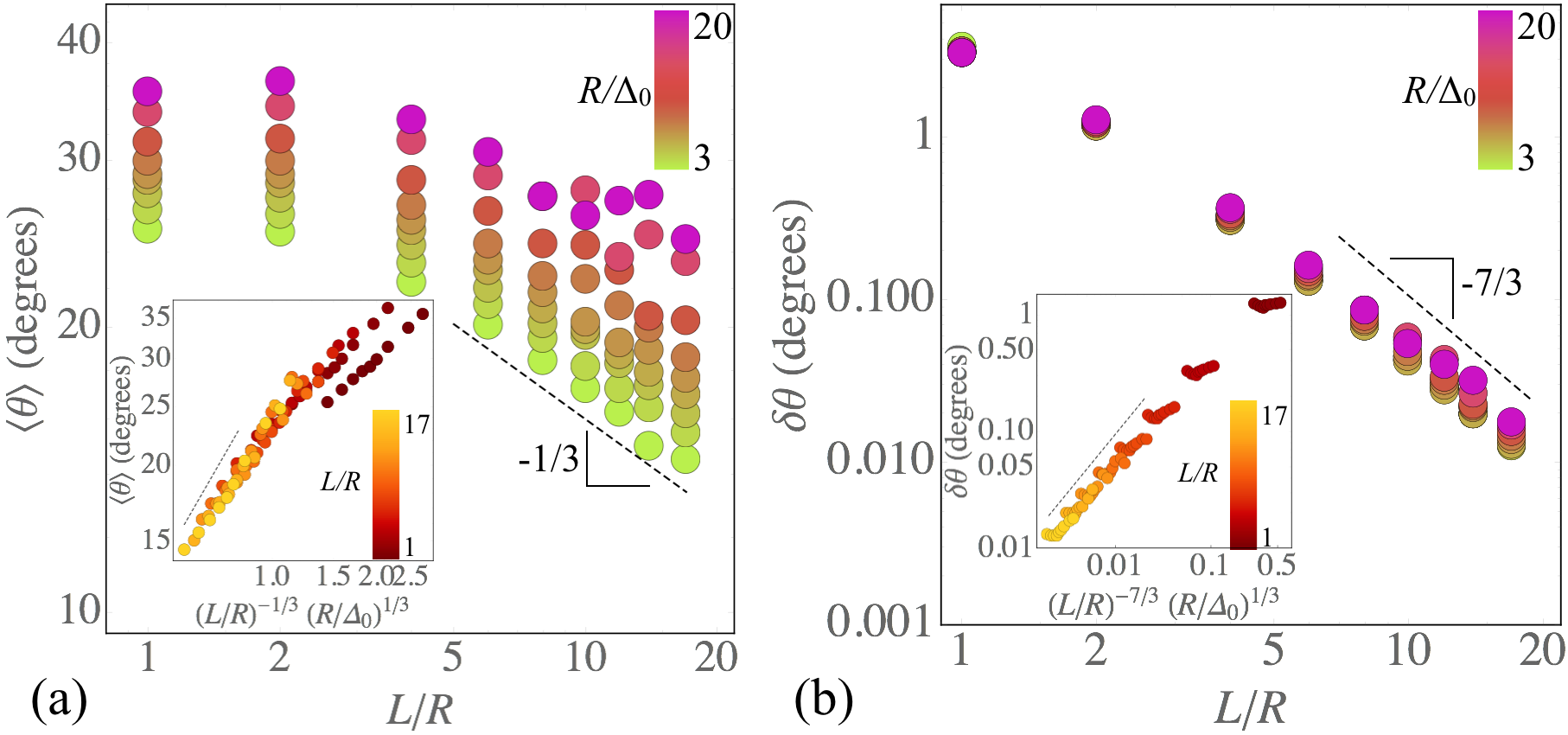}
\caption{Plots of (a) the mean tilt angle, and (b) the mean tilt angle deviation, vs.\ bundle aspect ratio $L/R$. Dotted lines show the power laws predicted from eqn (\ref{eq: scaling}), which follows the limit $R / \Delta_0 \gg 1$. Inset plots show the same data plotted vs.\ the dimensionless combinations given in eqn (\ref{eq: scaling}). Note that the double-asymptotic limit of $L/R \gg R/\Delta_0 \gg 1$ underlies eqn (\ref{eq: scaling}), highlighted by the yellow (lighter) data points in these insets.}
\label{fig:7FoldRodsData}
\end{figure}

\begin{figure*}[t!]
\centering
\includegraphics[width=0.98\textwidth]{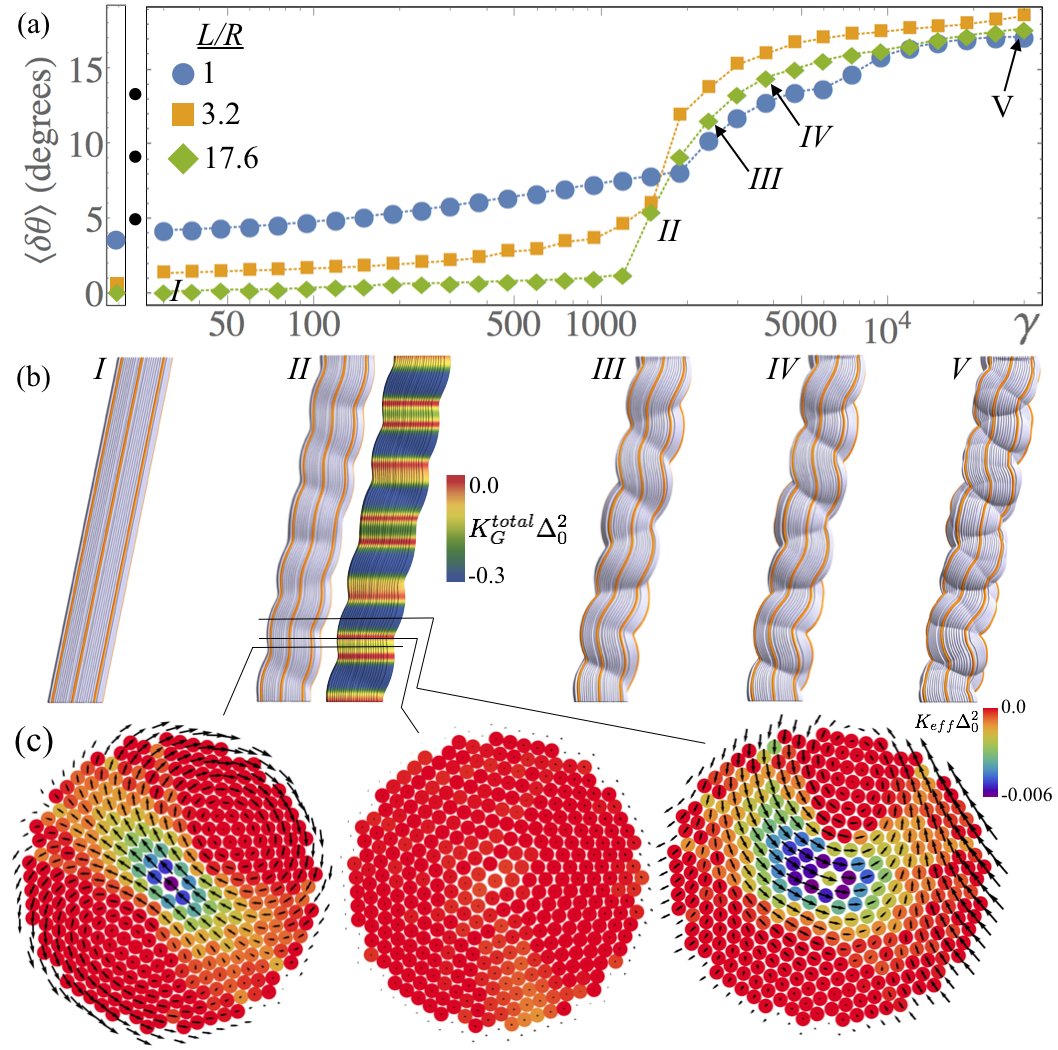}
\caption{Wrinkling instability for bundles with centered 7-fold disclinations. (a) The mean angle with respect to the bundle's centerline $\langle \delta \theta \rangle$ vs.\ $\gamma$, for various aspect ratios $L/R$. (b) Example structures for $L/R = 17.6$, with each structure, labeled {\em I-V} marked above in (a). Vertices near the ends are removed to highlight the bulk patterns. Structure {\em II} also has a representation coloring filaments by the total equivalent Gaussian curvature at each cross section along the {\em z} axis, calculated from eqn (\ref{eq:GaussForSurf}). (c) Select cross-sectional slices along the $L/R = 17.6$ bundle labeled {\em II}. Individual filaments are colored by their local Gaussian curvature $K_{\rm eff}$.}
\label{fig:7FoldSims}
\end{figure*}

Combining these energetic terms and minimizing with respect to mean-tilt and tilt-variation, we find
\begin{equation}
\label{eq: scaling}
\langle \theta \rangle \sim \Big(\frac{L}{R} \Big)^{-1/3} \Big( \frac{R}{ \Delta_0}\Big)^{1/3};\ \delta \theta \sim  \Big(\frac{L}{R} \Big)^{-7/3} \Big( \frac{R}{ \Delta_0}\Big)^{1/3}.
\end{equation}
While this scaling suggests that both angles vanish as $L \to \infty$, $\delta \theta$ (i.e.\ the double-vortex tilt pattern) is expected to decrease far more rapidly with bundle length. Fig.~\ref{fig:7FoldRodsData}(b) indeed shows this decrease with $L/R$, and is in further agreement with the weaker power-law dependence on $R/\Delta_0$. Fig.~\ref{fig:7FoldRodsData}(a) shows that the mean tilt value, $\langle \theta \rangle$, falls with both $L/R$ and $R/\Delta_0$, in numerical agreement with the power law of eqn.\ (\ref{eq: scaling}) in the asymptotic regime $L/R \gg R/\Delta_0 \gg 1$. This scaling suggest that such tilt patterns for shorter bundles of rigid filaments better screen defect stresses than longer ones.

We turn now to the case of finite flexibility (i.e.~$\gamma \neq 0$), and map out the $\gamma$-dependence of the 7-fold disclination buckling transition with the discrete filament bundle model. Results are found in a manner similar to the case of 5-fold bundles, but with $\gamma$ increased in a stepwise manner between energy minimizations in order to investigate the buckling transition. Final results are shown in Fig.~\ref{fig:7FoldSims}(a) for a bundle with $N_f = 428$ ($R \approx 10\Delta_0$) and three different aspect ratios $L/R$, where we measure the degree of shape buckling by $\langle \delta \theta \rangle$, the mean variation of filament tangents with respect to centerline.

For short bundles, we find a gradual increase of $\langle \delta \theta \rangle$ with $\gamma$, consistent with the intuitive notion that filament flexibility simply reduces the cost of the counter twist tilt pattern previously shown to be favorable for $\gamma \to 0$. As bundle length increases to the large aspect ratio limit, $L/R \gg 1$, we find that this gradual increase sharpens. For small $\gamma$, consistent with the scaling above, $\langle \delta \theta \rangle$ tends to zero with increasing $L/R$. However, we find that the large-$\gamma$ buckling persists and tends toward an $L$-independent behavior as $L/R$ increases. These trends are consistent with the emergence of a finite buckling threshold as $L\to\infty$, as predicted by the linear stability of the axisymmetric model in Section \ref{sec:linStab}. The behavior of the critical {\it fil}-vK number for this emergent double-vortex instability is schematically illustrated in the inset of Fig.\ \ref{fig:LinStabPred}(b). Compared to the rotationally symmetric constrained {\it ansatz} from the continuum elasticity stability analysis, this tilt pattern is extrapolated at the $L \rightarrow \infty$ limit to occur at a lower $\gamma_c \approx 1,500$ and $kR \approx 3$. As discussed, this finite-$\gamma$ threshold for buckling by 7-fold defects can be attribute to the elastic costs of breaking lengthwise symmetry of inter-filament spacing, a necessary consequence of the $K_{\rm eff} < 0$ tilt pattern.

Fig.~\ref{fig:7FoldSims}(a) shows the tendency with increased $L/R$ towards something like a second-order transition occurring around $\gamma \approx 2,000$. This value is notably an order of magnitude smaller than the threshold predicted by the continuum theory (from Fig.\ \ref{fig:LinStabPred}(b)). In large part, we expect this discrepancy in the buckling threshold is due to the fact that 7-fold bundles adopt shapes that are far from the axisymmetric ansatz of radial splay, and presumably, relax elastic costs far more efficiently. The specific pattern of buckling is comparable to the counter twist tilt pattern exhibited by finite-$L$ rigid filament bundles shown in Fig.~\ref{fig:7FoldRodsImage}. Additionally, we find that for large $L$, complex patterns are (at least for intermediate $\gamma$) consistent with a varying pattern of local counter twist along the bundle's length, where the span of one counter twist zone is much smaller that $L$. We can expect that the length of these counter twist zones, $h$, is set by a balance between inter- and intra-filament elasticity. We then balance over a vertical length $h$ the inter-filament elastic cost, $\sim Y (\delta \theta)^2 (h/R)^2$, with the bending cost, $\sim K (\delta \theta/h)^2$, to find an optimal size at $h_* \sim \gamma^{1/4} R = \sqrt{ \lambda_b R}$. This length is notably the same characteristic scale of optimal splay undulations. Beyond this length scale, the buildup of in-plane strains must be relaxed by a boundary layer that allows filament directions to reorient (e.g.~through a reversal of the counter twist tilt pattern).

The full 3D structures for various values of $\gamma$ can be seen in Fig.~\ref{fig:7FoldSims}(b); while Fig.~\ref{fig:7FoldSims}(c) shows maps of cross-sectional mean Gaussian curvature $K_{\rm eff}$ at distinct heights. For intermediate $\gamma$, the midsection of the bundle appears to alternate in the handedness of the double vortex, seen in Fig.~\ref{fig:7FoldSims}(c). Although the exact tilt pattern is not as simple as the ideal case found in Fig.~\ref{fig:7FoldRodsImage}(c), it clearly is not the splay ansatz proposed in Fig.\ref{fig:FirstLook}(d), and can be interpreted as a non-symmetric double vortex pattern. However, as $\gamma$ is increased still further, the buckled structures lose their alternating double vortex character and instead become more disorganized, possibly an indication that larger effects of inter-filament ``friction" at high curvatures in our model prevent full equilibration. Nevertheless, we note that these ``crumpled" shapes continue to show increasing negative effective curvature for larger $\gamma$, as well as a further decreasing distance between adjacent ``crumples", consistent with the expected decrease in the span of counter twist domains $h \sim \gamma^{-1/4}$. All in all, these shapes are far from the proposed radial undulations in Section \ref{sec:linStab}, a detail that will be discussed in the next section.

\section{Discussion}
\begin{figure}[t!]
\centering
\includegraphics[width=0.48\textwidth]{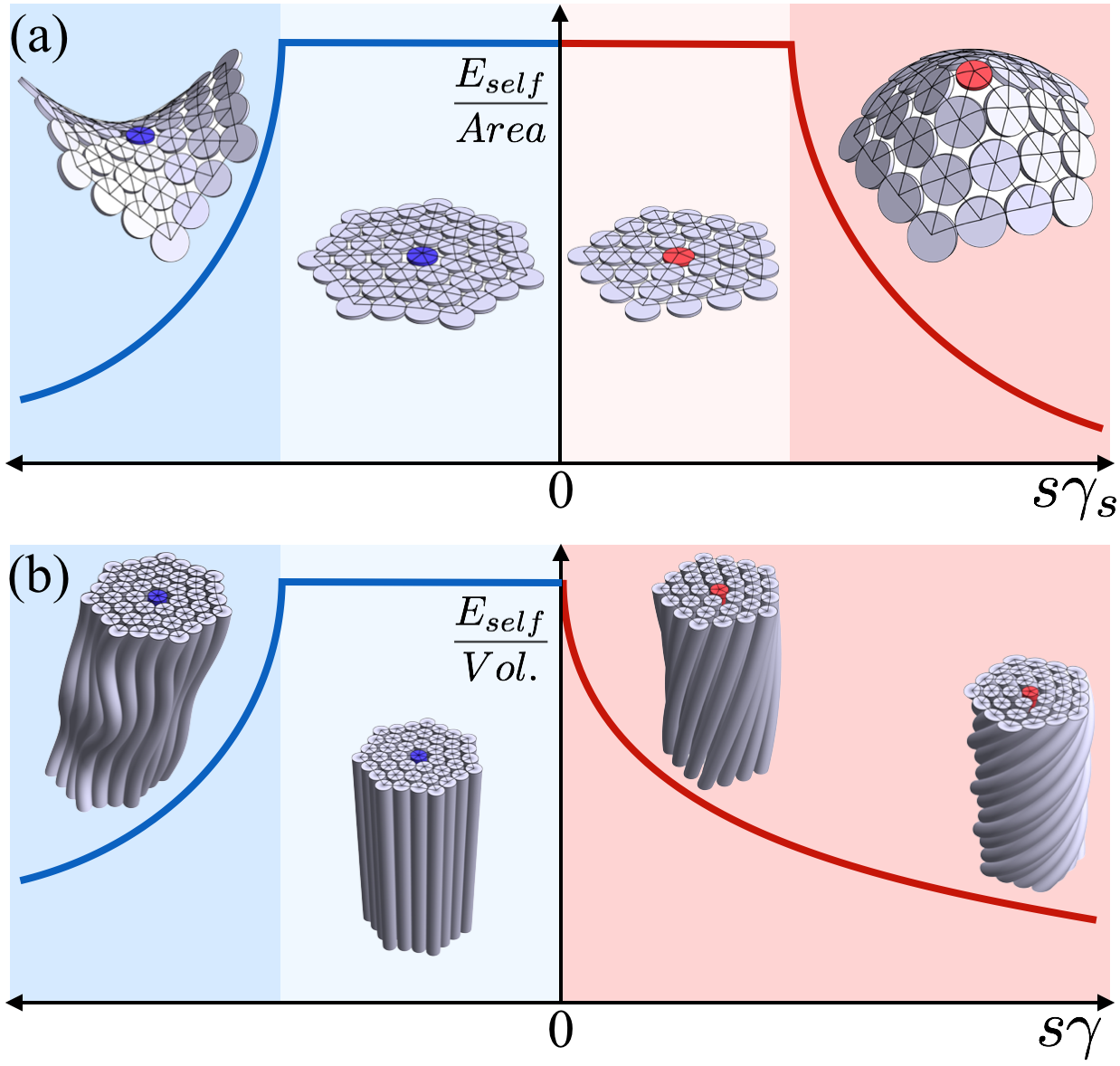}
\caption{(a) Self energy of a single defect vs.\ the disclination charge and FvK number, showing only a minor asymmetry in the threshold $\gamma_s$ for buckling. (b) Self energy of a single defect vs.\ the disclination charge and {\it fil}-vK number.}
\label{fig:selfEnergy}
\end{figure}

In the classical theories of disclinations, dating back to the geometric constructions of Volterra \cite{kleman} up to modern elastic theory treatments \cite{Romanov1981a, Romanov1981b}, positive vs.\ negative disclinations are symmetric with respect to their energetic costs in linear elastic theory. This asymmetry is only weakly broken in the 2D crystals, due to the unequal costs of conical- vs.\ saddle-like bending, and leads to few percent shift of the critical FvK number for buckling \cite{Seung1988a}.  Alternatively, disclination driven buckling in columnar structures appears to be in a distinct class, where the geometric packing constraints select one of the two signs of defects as particularly low energy, as shown schematically in the single defect self-energy plotted in Fig.~\ref{fig:selfEnergy}. Five-fold defects drive torsional buckling of bundles of any diameter and filament stiffness (i.e.~any nonzero $\gamma$), while for $L\to \infty$, 7-fold defects only drive shape transitions above a threshold bundle diameter (threshold $\gamma$). Simply put, this implies that positive disclinations should always be more abundant than negative disclinations in columnar/filamentous bundles. Here, we discuss the geometric origins of this asymmetry between defect signs, as well as the susceptibility to defect-induced instabilities for various material systems.
\begin{figure}[b]
\centering
\includegraphics[width=0.48\textwidth]{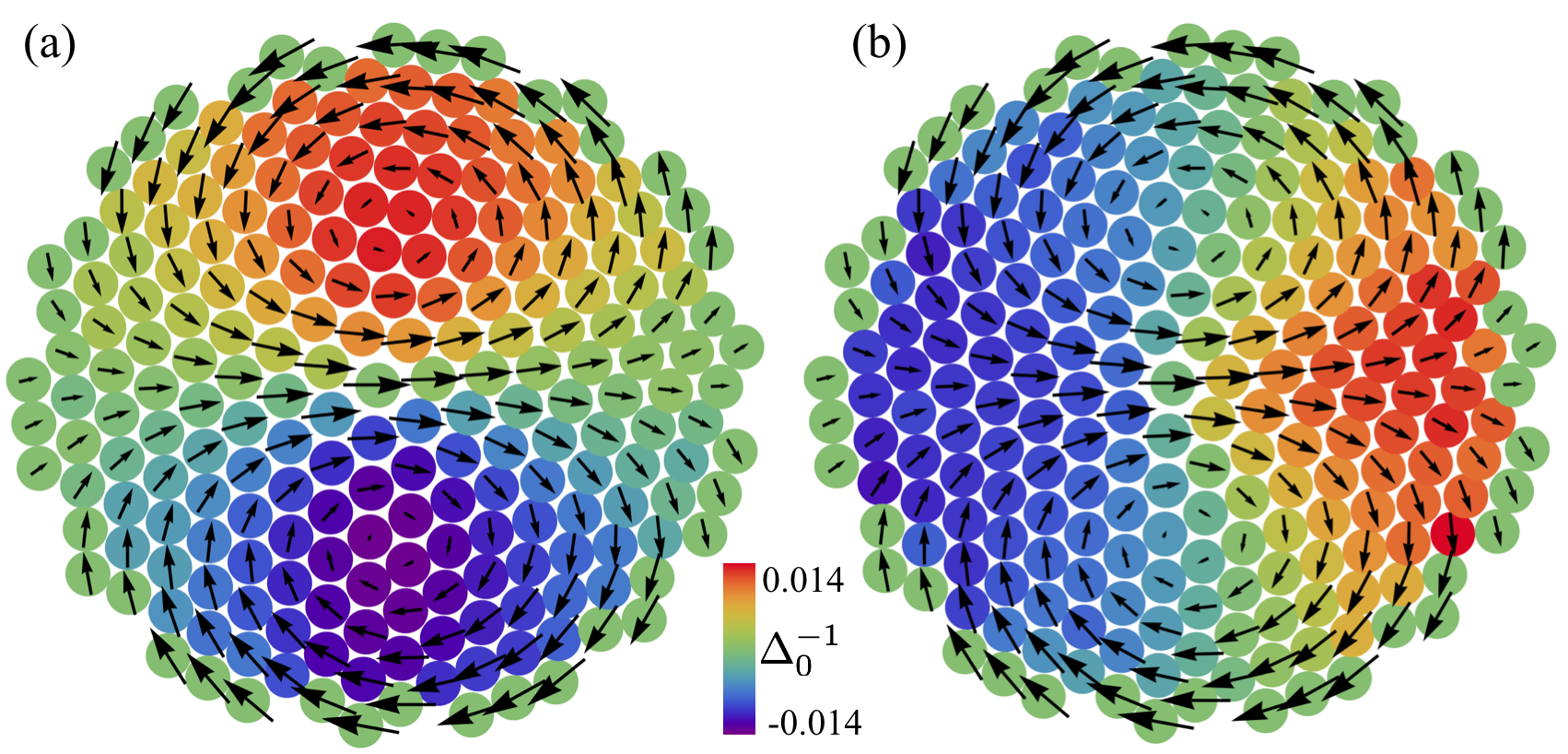}
\caption{Cross-sectional projections of infinitely rigid filaments in a 7-fold defective bundle exhibiting the double vortex tilt pattern. $R/\Delta_0 = 7$ and $L/R = 1.2$.(a) The double-twist, $\partial_x \tv_y - \partial_y \tv_x$, showing coordinated left- (blue) and right-handed (red) domains. (b) The splay-density of the strain variation, $v_{ii}$, showing a pattern orthogonal to the double-twist, where filament tilt causes a concentration of filaments on the left (blue) and depletion on the right (red).}
\label{fig:strainVar}
\end{figure}

\begin{figure*}[t!]
\centering
\includegraphics[width=0.98\textwidth]{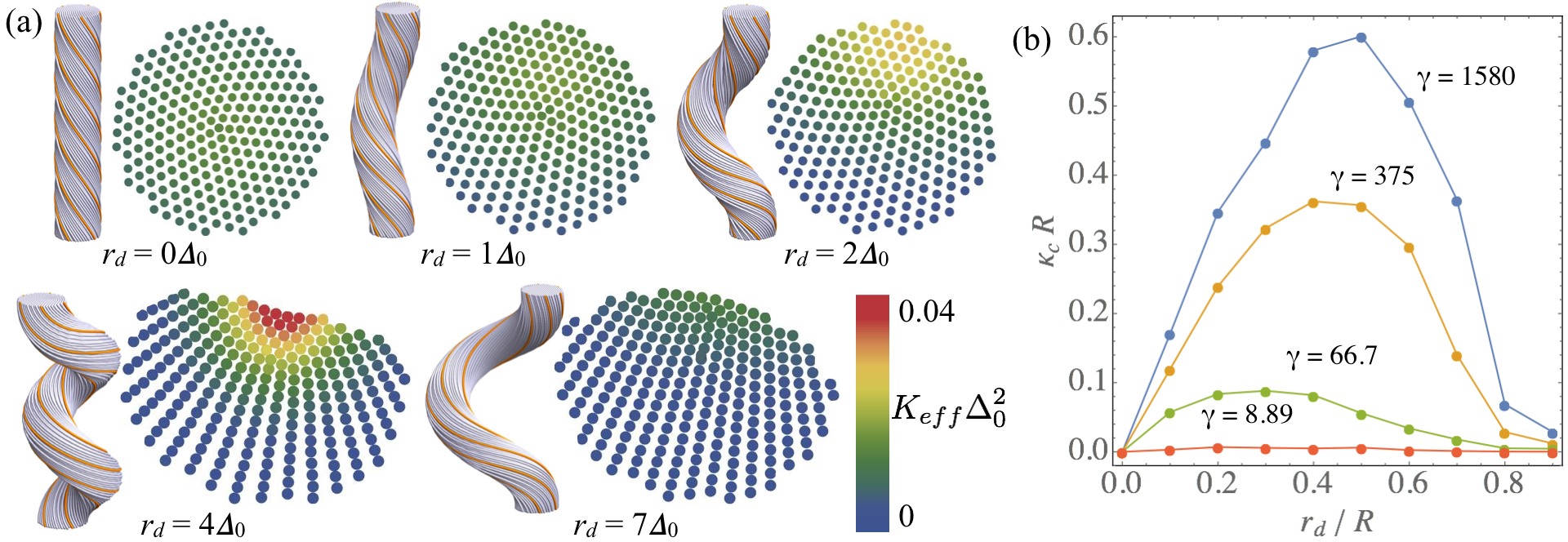}
\caption{(a) Visualizations of $R/\Delta_0 = 10$ bundles containing 5-fold disclinations displaced off-center at a distance $r_d$, and with $\gamma = 1,580$. Cross-sectional views show the equivalent Gaussian curvature, $K_{\rm eff}$. (b) The mean curvature of the centerline of a bundle, $\kappa_{c}$, as a function of the defect's position.}
\label{fig:offCentered}
\end{figure*}

Simulated elastic energy ground states for 7-fold defects reveal a surprising symmetry breaking feature, where buckling breaks both the lengthwise and axial symmetry of the initially parallel bundle (e.g.~Fig.~\ref{fig:7FoldRodsImage}). What is perhaps most surprising, is the emergence of the common tilt pattern of local {\it double-twist}, in response to both positive- and negative-charged disclinations. Double-twist is a cylindrically symmetric pattern of filament tilt, in which filament orientations twist around a central axis. A single double-twist domain has the tilt pattern $\tv_\perp \simeq \Omega r \hat{\phi}$ and a metric equivalent curvature $K_{\rm eff} = 3 \Omega^2$, which is consistent with the ability of this structure to screen positive disclination stresses. Why then, should 7-fold defects also drive the formation of double-twisted tilt patterns if they desire negative equivalent curvature?

The answer to this question derives from the interplay between the geometry of 2D tilt patterns in a given cross section and the lengthwise variation of inter-filament distances in the bundle. To illustrate this, we define the strain variation, $v_{ij} \equiv \partial_z u_{ij}$, to measure changes in inter-column strain along the bundle. From eqn (\ref{eq:ContElStrain}), we have
\begin{equation}
v_{ij} \simeq \frac{1}{2} \big[ \partial_i t_j + \partial_j t_i -\kappa (t_i n_j + n_i t_j) \big],
\end{equation}
where $\partial_z \tv \simeq \kappa {\bf n}$, with $\kappa$ and ${\bf n}$ being the respective local curvature and normal of a filament's Frenet-Serret frame. This tensor shows that certain tilt patterns in a 2D cross section require the buildup of elastic strains up- or down-stream from the section when $v_{ij} \neq 0$. Furthermore, it can be shown that uniform double-twist is, in fact, the only non-parallel tilt pattern which does not required lengthwise variations of inter-filament distances~\cite{Atkinson2017}. Because double-twist has strictly anti-symmetric in-plane gradients (i.e.~$\partial_i t_j = \epsilon_{ij} \Omega$), it is easily verified that all components of $v_{ij}$ vanish for a single, uniform-$\Omega$ domain. Again, the lack of lengthwise strain variation for uniform double-twist is linked to the absence of a threshold {\it fil}-vK for torsional buckling in 5-fold defective bundles.

For the case of 7-fold defective bundles, it is now easy to see why the axisymmetric radially splay {\it ansatz} is not preferred. Although it does generate a favorable local distribution of $K_{\rm eff} <0$, this tilt pattern unfavorably requires $v_{ii} \simeq \grad_\perp \cdot \tv_\perp \neq 0$ everywhere in the cross section. Therefore, this axisymmetric geometry is presumably a poor compromise between the preference for negative $K_{\rm eff}$ and minimal elastic strain variation along the bundle. Alternatively, the counter-twist observed in Section \ref{sec:7foldSim}, composed of two oppositely rotating double-twisted domains, focuses a region of highly-negative curvature between the opposing domains (near to the disclination position). While simultaneously, a minimal lengthwise strain build-up along the bundle cross section is maintained. This is illustrated in Fig.~\ref{fig:strainVar}, which shows the map of double-twist and splay-density ($v_{ii}$), for a 2D cross section of a simulated 7-fold defective bundle of rigid filaments found in Section \ref{sec:7foldSim}. Unlike radial splay, the counter-twist tilt pattern effectively expels splay from most of the bundle's cross section, while generating a sufficient measure of $K_{\rm eff}<0$ (Fig.~\ref{fig:7FoldRodsImage}(c)) near the bundle's center to screen the in-plane stresses generated by the 7-fold defect.
\begin{table*}
\begin{tabular}{l|c|c|c|c|c||c}
Filament Type & Filament & Bundle & Bending & Cohesion & Cohesive & {\it fil}-vK, $\gamma$ \\
& Diameter, $d$ & Radius, $R$ & Stiffness, $B$ &(per length), $\epsilon$ & Range, $\sigma$ & \\
\hline
\hline
carbon nanotubes \cite{Volkov2010, Wang2015d, Thess1996, Zhang2004} & $ 1.2 : 2.7~{\rm nm}$ & $ 4 : 500~{\rm nm} $ & $ 0.2 : 25~{\rm keV~nm} $ & $ 5 : 10~{\rm eV / nm} $ & 0.3 : 0.5~{\rm nm} & $ 10^{-6} : 10^{4} $ \\ \hline

DNA \cite{Leforestier2009a, Hud2001a, Qiu2011b, He2013, Bloomfield1997} & $ 2~{\rm nm} $ & $ 10 : 30~{\rm nm} $ & $\sim 58~k_B T~{\rm nm}$ & $ 0.1 : 2.5~k_B T / {\rm nm}$ & $ 1 : 3~{\rm nm}$ & $ 0.001 : 1 $ \\ \hline

microtubules \cite{Gittes1993, Needleman2004a, Needleman2004, Hilitski2015, Safinya2016} & $0.025~{\rm \mu m}$ & $ 0.05 : 1~{\rm \mu m}$ & $\sim 5200~k_B T~{\rm \mu m}$ & $ 10 : 200~k_B T / {\rm \mu m}$ & $ 0.5 : 10 ~{\rm nm}$ &  $ 10^{-8} : 0.1$ \\ \hline

microfibers/posts~\cite{Pokroy2009a, Bico2004a} & $ 0.2 : 100 ~{\rm \mu m}$ & $ 2 : 200 ~{\rm \mu m}$ & $ 8\mathrm{e}{-6} : 5\times 10^{5} ~{\rm nJ~\mu m}$ & $ 10^{-5} : 10^{-2}~{\rm nJ /\mu m}$ & $ 0.1 : 100 ~{\rm \mu m} $ & $ 10^{-9} : 10^{6} $

\end{tabular}
\caption {Estimated values of {\it fil}-vK for cohesive filament bundles of different materials. Microtubule bundles parameters have been estimated considering interactions that range from relatively weak and long-range depletion~\cite{Needleman2004, Hilitski2015} to shorter-range and stronger binding due to polyvalent counterion binding~\cite{Needleman2004a, Safinya2016}.  Here, cohesion between polymer microfiber arrays is modeled by capillary bridging~\cite{Pokroy2009a, Bico2004a}.}
\label{tab:morph}
\end{table*}

The above arguments suggest that the unique geometry of double-twist make this tilt pattern a potentially inexpensive ``building block" for more complex 3D buckled cohesive bundles, well beyond the seemingly ideal centered 5-fold defects. The generic emergence of double-twist can further be illustrated by the shape equilibria of bundles possessing off-centered 5-fold disclinations, shown in Fig.~\ref{fig:offCentered}. Here we see that as the defect position, $r_d$, increases away from the bundle center, the centerline of the bundle buckles helically with a curvature that increases with $\gamma$. The equilibrium curvature of the centerline, $\kappa_c$, depends non-monotonically on $r_d$. Superficially, this buckling pattern would seem to imply a qualitatively different responses of non-axisymmetry defect distributions. On the contrary, this ``writhing" bundle equilibria are in fact themselves regions of a uniform double-twist domain, but with the center of rotation located away from the bundle center and eventually outside of the bundle's cross section. This behavior is likely the result of non-linear (large-tilt) corrections to the metric geometry of uniform double-twist described previously \cite{bruss2012c}. Here $K_{\rm eff}$ is concentrated at at the center of rotation and tapers to an effectively flat geometry for large radial distances from this axis (compared to helical pitch). Thus, the shifting center of double-twist rotation is likely driven by the polarized stress distribution created by off-center defects. The bundles ultimately straighten for larger $r_d$ due to the elastic screening of defect stress by the free boundary \cite{grason2010a, grason2012c}.

Next, we consider the relevance of defect-induced shape buckling in cohesive filament bundles realized for a range of physical systems. In particular, we estimate the range of accessible {\it fil}-vK numbers based on measured or predicted values of their resistance to bending, their elasticity of inter-filament cohesion, and their observed values of bundle radius, $R$. The buckling behavior to defects is critically sensitive to $\gamma \equiv (R/\lambda_b)^2$, which is the ratio of bundle size relative to the material dependent length scale, $\lambda_b = \sqrt{K/Y}$, itself a measure of the ratio of intra-filament to inter-filament elasticity. The bend elastic cost can be estimated simply from the bending stiffness of filaments $B$, and their diameters $d$, as $K \approx B/d^2$. Far less well characterized is the elasticity of inter filament cohesion, which in turn, controls the elasticity of the 2D filament array. Following refs \cite{bruss2013a, Hall2016}, we estimate $Y$ by assuming that the elastic stiffness of inter-filament cohesion (strictly speaking the curvature of the inter-filament binding potential) can be estimate as $\epsilon/\sigma^2$, where $\epsilon$ is the cohesive energy per unit length between neighbor filaments, and $\sigma$ is a length scale characterizing the cohesive range. Neglecting numerical prefactors accounting for lattice geometry, this estimate then gives $Y \approx \epsilon/\sigma^2$.

In Table \ref{tab:morph}, we compare values of $\gamma$ for vastly different classes of cohesive bundles: carbon nanotubes; biological filaments (DNA and microtubules); and polymeric microfibers assembled by surface (or capillary) interactions. Significantly, we find that the materials can reach a $\gamma$ that spans up to 15 orders of magnitude, up to $\gamma \sim 10^{6}$. In general, this suggests that almost all cohesive filament bundle assemblies are susceptible to buckling by 5-fold disclinations, but in general, most conditions fall below the range where they are perturbed by 7-fold disclinations (i.e.~$\gamma \gtrsim 2000$).  For example, nanotube bundles that condensed from solutions~\cite{Thess1996} grow to relative narrow widths ($\sim 100~{\rm nm}$) and reach only $\gamma \sim 10^{-6}-10^{-5}$. Alternatively, much larger  ($\sim 5 ~{\rm \mu m}$) ``yarns" generated by spinning from nanotube ``forests" extend far into the large {\it fil}-vK range, potentially exceeding the critical $\gamma$ for the 7-fold disclination-induced shape buckling shown in Fig.~\ref{fig:7FoldSims}.

Microtubule bundles formed by polyvalent counterion condensation have been observed to exhibit long-wavelength undulations, interpreted as 3D writhing configurations \cite{Needleman2004a}, not unlike the helical buckling of off-center defect patterns of Fig.~\ref{fig:offCentered}. For these systems there is also some evidence of ``irregular" packing in the 2D cross section, though at present there has been no attempt to quantify defect distributions in these experiments. It has been proposed that a non-equilibrium process of cohesive condensation of filament bundles may often lead to the trapping of topological defects into the cross-sectional order~\cite{Gov2008}, which may have consequences on their structure and assembly. From the values of interaction and mechanics of microtubules, in particular for relatively brittle binding by polyvalent counterion bridging, we modestly estimate that $\gamma \sim 0.1$. This small allowance for flexibility may drive a detectable degree of helical buckling in the presence of positive disclinations, as suggested by the low-$\gamma$ range of Fig.~\ref{fig:5foldcurvconc}.

Finally, we conclude with a simple estimate for macroscopic and elastic filaments, held together by cohesive surface contact (i.e.~the range of attractions is small compared to the filament diameter). In this case, both filament bending stiffness, $B \approx E_f d^{4}$, and inter-filament contact stiffness, $Y \approx E_f$, are proportional to the elastic modulus, $E_f$, of the filaments. Hence, their ratio is expected to be {\it independent} of $E_f$, and therefore roughly $\lambda_b \approx d$. Because $R \approx d N_f^{1/2}$ in this strong cohesive contact regime, we expect that the {\it fil}-vK becomes {\it independent} of material parameters and simply dependent on the number of filaments within the cross section, $\gamma \approx N_f$.

\section{Conclusion}
We have determined the disclination-induced buckling instabilities for cohesive filament bundles, and resolved their dependence on the size and mechanics of the assemblies. These results point to the ratio of intra-filament to inter-filament elastic stiffness as the key material-dependent quantity that regulates the emergent shapes of defective bundles. Compared to flexible membranes, the equilibrium shapes of bundles exhibit a far greater non-trivial dependence on the types of defects. This results from the interplay between the local metric geometry of 2D bundle cross sections and the lengthwise variations of filament spacing within the bundle. We find that bundles are vastly more susceptible to shape deformation by 5-fold defects (positive disclinations) than 7-folds. While bundles possessing other cross-sectional symmetries (e.g.\ 4-fold) may require anisotropic elastic costs as well as distinct topological charges of defects, we expect this basic conclusion to hold independent of lattice symmetry, as the geometric principles underlying metric coupling to tilt and uniform longitudinal spacing are generic.

The principles developed in this study suggest new strategies for engineering the 3D shapes of bundles through their controlled 2D packing. We envision various synthetic strategies that may be exploited to template columnar or filamentous assemblies with controlled topological defects in their cross-sectional order. This feat would be much in the same spirit of kirigami engineering of point-like disclinations used to engineer the 3D (and self-folding) shapes of sheets. This could be achieved, for example, by i) the capillary cohesion of nanofabricated 2D arrays of high aspect ratio flexible pillars \cite{Pokroy2009a}, ii) the controlled defect formation in columnar assemblies grown epitaxially from a lithographically templated surfaces \cite{Xu2011, Bosse2007}, and iii) so-called DNA-origami techniques to engineer dsDNA bundles \cite{Deitz2009} with programmed 2D cross sections, containing intentionally placed disclinations and dislocations. For example, as shown in Fig.~\ref{fig:offCentered}, controlling the placement of a 5-fold defect and lead to spontaneously writhing and twisting 3D configurations, dramatically reshaping the responses of the assembly to a range of stimuli (e.g.\ mechanical, electronic, photonic, etc.).   In systems where the cohesion between filaments and their elastic stiffness may be externally tunable (say in temperature- or field-responsive materials), we envision that the 3D shapes of such bundles could be adjustable, driven to wind and unwind responsively.

To a first approximation it can be anticipated that multiple elementary disclinations of the same charge lead to effects equivalent to single higher charged defects, namely, added drive for positive or negative curvature bundle textures. However, given the highly non-linear dependence of buckling to disclination type, the response to even a single edge {\it dislocation}, i.e.\ a 5-7 ``dipole", is likely to be far more complex for bundles than for their geometrical analogs of 2D membranes \cite{Seung1988a}. Previous work provides some insight into how dislocations drive certain tilt patterns \cite{Grason2012d}. First off, dislocations have an extra degree of motion over disclinations, namely their orientation. In much the same way 5-fold disclinations are expected to be more prevalent in filament bundles, dislocations are expected to be oriented such that the 5-fold end is turned towards the center of twist (corresponding to the removal of a partial row extending radially to the boundary), this generate stresses that can be relaxed by positive curvature tilt patterns (e.g.\ twist). Furthermore, the strength of the tilt-dislocation stress coupling is strongly position dependent, with a preference for radially-aligned dislocations situated at $R/\sqrt{3}$ from the bundle center. Alternatively, misoriented dislocations (e.g.\ with the negative disclination end closer to the center) cannot be relaxed by twist, and may instead lead to non-axisymmetric and longitudinally asymmetric deformations modes. Lastly, unlike disclinations, whose strength is quantized in multiples of $\pi/3$, the net strength of dislocations to drive shape instabilities depends on the ratio of Burger's vector (a microscopic dimension proportional to lattice spacing) to bundle size, $b/R$. Therefore, we would anticipate that the drive to buckle the bundle into a 3D shape may be more sensitively controlled by the collective effects (number, orientation, and locations) of multiple dislocations arranged in a bundle as compared to one or a few isolated dislocations.

While such experimental directions remain to be explored, the principles governing the response to elementary 5- and 7-fold disclinations lay the ground work for engineering custom-made defect configurations in 2D filament arrays, which could be used to ``program" specified 3D shapes of bundles. However, beyond the isolated disclinations discussed here, the geometric and mechanical principles that govern the collective responses to multi-defect patterns remain open.

\begin{acknowledgments}
We would like to thank A. Azadi for useful discussions as well as D. Hall and D. Atkinson for valuable input on this manuscript. This work was supported by the National Science Foundation through Award No. DMR 16-08862.  We would also like to acknowledge the hospitality of the Aspen Center for Physics (supported by NSF through Award No. PHY 16-07611) where some of this work was completed.  Numerical computations were performed on the UMass Shared Cluster at the Massachusetts Green High Performance Computing Center.
\end{acknowledgments}

\appendix
\section{Continuum and discrete model parameters}
\label{sec:ContinuumVsDiscrete}
Here we describe the correspondence between parameters of the discrete filament model of Sec.~\ref{sec:ModelIntro} and the elastic moduli of the continuum theory of columnar bundles presented in Sec.~\ref{sec: continuum}. We begin with the squared curvature of the $i$th filament with tangent ${\bf T}_i (s)$
\begin{equation}
\kappa^2 = \left( \frac{\partial {\bf T}_i}{\partial s} \right)^2 \simeq \frac{({\bf T}_{i,n+1} - {\bf T}_{i,n} )^2}{\ell_n^2} ,
\label{eq:DiscCurvDef}
\end{equation}
where $\lim_{\ell_n \to 0}$ is the continuum limit. We can convert from the discrete model of bending energy in eqn (\ref{eq:CoarseBend}), to the continuum model in eqn (\ref{eq:ContElBend}) by summing over all ``bending bonds" on filament $i$,
\begin{align}
E_{b}^{(i)} &= \frac{B}{2} \int_0^L ds~ \kappa^2(s) 
&\simeq \frac{B}{\ell_n} \sum_{n=1}^{N_v-1}(1 - {\bf T}_{i,n}  \cdot {\bf T}_{i,n+1}),
\end{align}
and then summing over all filaments, yielding the total bending energy
\begin{equation}
\sum_{i=1}^{N_f} E_{b}^{(i)} \simeq \frac{K}{2} \int dV ~ \kappa^2(\xv) ,
\end{equation}
where $\int dA \simeq  \sum_{i=1}^{N_f} \rho_0^{-1}$, and $\rho_0 =\Delta_0^{-2}/\sqrt{3}$ is the stress-free areal density of filaments giving,
\begin{equation}
\label{eq:BendConversion}
K = \frac{ 2 B}{\sqrt{3} \Delta_0} .
\end{equation}

For the discrete model of the inter-filament elasticity, from eqn (\ref{eq:CoarseCohesive}) we have the elastic energy of the $n$th layer,
\begin{equation}
E^{(n)}_{elas}= \frac{\epsilon}{2} \sum_{i=1}^{N_f} \sum_{\langle ij \rangle} (\Delta_{ij} - \Delta_0)^2,
\label{eq:CohDiscrete}
\end{equation}
where $\Delta_0$ is the preferred local spacing between filaments. Structurally, the bundle model is essentially $N_v$ stacks of the hexagonal bead-spring model of ref.~\cite{Seung1988a}, with a geometrically non-linear coupling between local tilt and lattice strain. Thus, summing over these layers we have
\begin{equation}
\sum_{n=1}^{N_v}E^{(n)}_{elas} \simeq \frac{\sqrt{3} \epsilon}{4 \ell_0} \int_0^L  dz \int  dA ( u_{kk}^2+2u_{ij}^2) ,
\end{equation}
from which we have the elastic constants,
\begin{equation}
\lambda = \mu = \frac{\sqrt{3} \epsilon}{4 \ell_0} 
\end{equation}
corresponding to 2D Youngs modulus,
\begin{equation}
Y = \frac{4 \epsilon}{\sqrt{3} \ell_0},
\label{eq:CohConversion}
\end{equation}
and 2D Poisson ratio $\nu = 1/3$.

From these the {\it fil}-vK number may be estimated from the discrete model parameters as
\begin{equation}
\gamma = \frac{Y R^2}{K} \simeq \frac{2 \epsilon R^2 \Delta_0^2}{B \ell_0}.
\label{eq:FvKConversion}
\end{equation}

\section{Continuum elasticity stability analysis}
\label{sec:LinearStab}
We analyze the stability of lengthwise periodic and axisymmetric deformation patterns of bundles of radius $R$, with a centered disclination of charge $s$. To determine the equations of equilibrium, we begin with an initial displacement field ${\bf u} ( {\bf x})$, subject to a small perturbation $\delta{\bf u} ( {\bf x})$, and consider the variation of the energy, $\delta E = E[{\bf u} ( {\bf x}) + \delta{\bf u} ( {\bf x})]  - E[{\bf u} ( {\bf x})]$. Solving for the equations of equilibrium, we arrive at
\begin{align}
\partial_j \sigma_{ij} - \partial_z (t_j \sigma_{ij}) - K \partial_z^3 t_i &= 0 \text{~~~(force balance)} \label{eq:PartFoceBal}\\
dS_i \sigma_{ij} &= 0 \text{~~~(stress free sides)} \\
t_j \sigma_{ij} + K \partial_z^2 t_i &= 0 \text{~~~(stress free ends)} \label{eq:StressEnds}\\
K \partial_z {\bf t} =& ~0 \text{~~~(torque free ends)}.
\label{eq:EquilEqColumns}
\end{align}

We determine the conditions for the solutions to the stability equations outlined in eqns (\ref{eq:PartFoceBal})-(\ref{eq:EquilEqColumns}), by considering solutions that are weakly perturbed from the parallel state, and of the form
\begin{equation}
{\bf u}({\bf x}) = {\bf u}_0 ({\bf x}) + \epsilon {\bf u}_1 ({\bf x}) + \epsilon^2 {\bf u}_2 ({\bf x}) + \epsilon^3 {\bf u}_3 ({\bf x}) + \text{\ldots},
\label{eq:PertDisplacement}
\end{equation}
where $\epsilon$ is the amplitude of the deformation, taken to be arbitrarily small near the point of linear instability (i.e.~the supercritical bifurcation point), and ${\bf u}_n$ represents the $O(\epsilon^n)$ deformation modes. For linear stability, it is sufficient to analyze only the lowest order in $\epsilon$, though if we want to solve for the dependence on the $\epsilon$ distance from the instability, we need to solve to order $\epsilon^3$.

Considering only the order $\epsilon^1$ term, and decomposing the displacement into radial and azimuthal components
\begin{equation}
{\bf u}_1 ({\bf x}) = \rho({\bf x}) \hat{r} + \tau ({\bf x}) \hat{\theta}.
\label{eq:RadAzComp}
\end{equation}
Note that the functions $\tau(\xv)$ and $\rho(\xv)$ should not be confused with the dimensionless constants $\tau_0$ and $\rho_0$ that parameterize the amplitude of the twist and splay ansatz in Sec.~\ref{sec:linStab}. Applying this form to eqn (\ref{eq:PartFoceBal}) and assuming only axisymmetric patterns of deformation, we find the force balance along the $\hat{r}$ direction
\begin{equation}
(\lambda+2\mu)\partial [r^{-1} \partial_r (r \rho) ] - \sigma^0_{rr} \partial_z^2 \rho - K \partial_z^4 \rho = 0,
\label{eq:ForceBalRad}
\end{equation}
and along the $\hat{\theta}$ direction
\begin{equation}
\mu\partial [r^{-1} \partial_r (r \tau) ] - \sigma^0_{rr} \partial_z^2 \tau - K \partial_z^4 \tau = 0.
\label{eq:ForceBalAz}
\end{equation}
The boundary conditions on the sides of the bundle are simply $\sigma^1_{rr} = \sigma^1_{r\theta} = 0$, or specifically
\begin{align}
\lambda \rho(R)/R = (\lambda + 2\mu)\partial_r \rho(R) &= 0 \\
\partial_r \tau(R) - \tau(R)/R &= 0.
\label{eq:BoundConSigmas}
\end{align}
And finally, we have the boundary conditions for the derivatives of the displacements at the ends of the bundles, but we will neglect these by assuming that solutions are periodic, and of the form
\begin{equation}
\rho({\bf x}) =  \delta u_r (r) \cos(k z);~~ \tau({\bf x}) = \delta u_\phi \cos(k z).
\label{eq:RhoAndTauGeneral}
\end{equation}

To compare to a finite length bundle, we might consider wavelengths that are commensurate with the bundle length, $k = 2\pi n/L$, though to be clear, these purely sinusoidal deformations will not allow us to match the free end boundary conditions from eqns (\ref{eq:StressEnds}) and (\ref{eq:EquilEqColumns}). Presumably, a boundary layer is required to match the purely periodic solutions we consider, to the free end calculations. Therefore, we work under the assumption that the length scale of this boundary layer will vanish proportional to $\sqrt{K/Y}$, and hence can be ignored for large aspect ratio bundles ($L/R \gg 1$), and large bundle {\it fil}-vK number ($\gamma \gg 1$).

To proceed, we rewrite the equations in dimensionless variables, by measuring all lengths in units of bundle width $R$, and stresses in units of $Y$. Doing this, and recalling the definition of the 2D Poisson ratio, $\nu = \lambda/(\lambda + 2\mu)$, we rewrite eqns (\ref{eq:ForceBalRad}) and (\ref{eq:ForceBalAz}) in the form of eqn (\ref{eq: coulomb}) with the effective ``potentials"
\begin{align}
V_r(r) &= - \frac{s (k R)^2 (1-\nu^2)}{8 \pi} \ln (r/R) \\
V_\phi(r) &= - \frac{s (k R)^2(1+\nu)}{4 \pi} \big[ \ln (r/R) +1\big]
\label{eq: potential}
\end{align}
and eigenvalues
\begin{align}
\epsilon_r &= \frac{(1-\nu^2) (k R)^4}{2\gamma} \label{eq:SplayEig}\\
\epsilon_\phi &= \frac{(1+\nu)  (k R)^4}{\gamma}.
\label{eq: eigen}
\end{align}
In this way, we have recast the linear stability calculation in terms of an eigenvalue problem, with the boundary conditions
\begin{align}
\nu  \delta u_r(R) + \partial_r \delta u_r(R) &= 0 \\
\partial_r  \delta u_\phi (1) -  \delta u_\phi (1)/R &= 0.
\label{eq:BoundDim}
\end{align}
For the linear stability calculation, we are interested in the ground state solution, i.e.~the smallest values of $\epsilon_r $ or $\epsilon_\phi$ that are consistent with our boundary conditions. This will correspond to the first instability---the lowest value of $\gamma$---at which a given periodic mode $k$ becomes unstable. In the main text and Fig.~\ref{fig:LinStabPred}, we solve these equations for the most unstable wavenumber, $k$, given centered 5-fold ($s = +2\pi/6$) and 7-fold ($s = -2\pi/6$) disclinations.

\section{Discrete approximation of equivalent Gaussian curvature}
\label{sec:EquivKG}
For an arbitrary discrete surface composed of vertices, edges, and triangular faces, the Gaussian curvature can be defined in terms of the deficit of internal angle at a single vertex, as
\begin{equation}
K_G = 3 \left( 2 \pi - \sum_{\alpha=1}^{f} \psi_\alpha \right) /A,
\label{eq:GaussForSurf}
\end{equation}
where $\psi_\alpha$ is the internal vertex angle for face $\alpha$, $A$ is the summed area of all the faces attributed to the vertex, and $f$ is the number of triangular faces adjoining the vertex \cite{Meyer2003, Meek2000a}.  We can generate an equivalent mesh for a bundle by finding the points by intersection of all filaments with a plane, here chosen to be the $z$ plane; followed by a Delaunay triangulation. While this mesh is itself flat, with zero Gaussian curvature, the relevant distance between filaments is the distance of closest contact which is generally out of plane (i.e.~similar to $\Delta_{n,ij}$, but defined within a cross-sectional plane rather than at vertex $n$). Explicitly, if the in-plane spacing between two filaments, $i$ and $j$, is ${\bf \Delta} = {\bf x}_j - {\bf x}_i$, then the out of plane distance of closest contact from filament $i$ to filament $j$, is defined as
\begin{equation}
 {\bf \Delta}_\perp = {\bf \Delta} -{\bf t}_j ( {\bf \Delta} \cdot {\bf t}_j).
\label{eq:DoCCVec}
\end{equation}
Therefore, by computing the corrected side length, $|{\bf \Delta}_\perp|$ of a given triangle, we calculate the internal angles, $\psi_\alpha$.   From this we compute the effective Gaussian curvature, $K_{\rm eff}$, at given filament (and at a given $z$) by summing the internal angles in formula eqn (\ref{eq:GaussForSurf}) for the local triangular neighbor array. In essence, this method works by distorting the dimensions of every triangle based on the relevant distances of closest contact, then stitching them back together in a manner that preserves the original network topology, but requiring out-of-plane orientations. The total effective curvature of the surface, $K_{\rm eff}^{total}$, is simply the sum of the equivalent Gaussian curvature of each individual filament at that $z$ plane.

Although it is possible to always calculate the equivalent Gaussian curvature for a given pattern of filament tilt, it is not always possible to find an example of the equivalent surface that is embeddable in $R^3$, or one that is strictly unique, possessing the same metric data \cite{Abate2012}. The reconstructed surfaces in Fig.~\ref{fig:5foldcurvconc}(d) were found by energetically relaxing the positions of a bead-spring model of vertices, with a spring constant $k_s$ and a spring rest length of ${\bf \Delta}_\perp$ (which is unique for each vertex). Energy minimization was performed on an initially axisymmetric conical surface, and proceeded until the total energy of all the springs fell below $0.001 k_s$.

\bibliography{apstemplate}

\end{document}